\documentclass[preprint]{aastex61}

\usepackage{subfigure}
\usepackage{color}

\newcommand{\HH}{H$_{2}$}

\newcommand{\mum}{\ensuremath{\mu \mathrm{m}}}
\newcommand{\FeII}{[\ion{Fe}{2}]}

\newcommand{\HI}{H\,\textsc{i}}
\newcommand{\HII}{\ion{H}{2}}

\newcommand{\kms}{km s$^{-1}$}

\newcommand{\Lya}{Ly$\alpha$}
\newcommand{\redchi}{$\chi^2_{\nu}$}

\begin{document}

\title{The Orion fingers: H$_2$ temperatures and excitation in an explosive outflow
        }

\author{Allison Youngblood}
\altaffiliation{Now located at NASA Goddard Space Flight Center, Greenbelt, MD 20771, USA}
\affiliation{Laboratory for Atmospheric and Space Physics, University of Colorado, 600 UCB, Boulder, CO 80309, USA}
\affiliation{Department of Astrophysical and Planetary Sciences, University of Colorado, UCB 389, Boulder, CO 80309, USA}
\email{allison.a.youngblood@nasa.gov}

\author{Kevin France}
\affiliation{Laboratory for Atmospheric and Space Physics, University of Colorado, 600 UCB, Boulder, CO 80309, USA}
\affiliation{Department of Astrophysical and Planetary Sciences, University of Colorado, UCB 389, Boulder, CO 80309, USA}
\affiliation{Center for Astrophysics and Space Astronomy, University of Colorado, 389 UCB, Boulder, CO 80309, USA}

\author{Adam Ginsburg}
\affiliation{National Radio Astronomy Observatory, Socorro, NM 87801, USA}

\author{Keri Hoadley}
\affiliation{Department of Astronomy, California Institute of Technology, 1200 East California Boulevard, Pasadena, CA 91125, USA}

\author{John Bally}
\affiliation{Department of Astrophysical and Planetary Sciences, University of Colorado, UCB 389, Boulder, CO 80309, USA}
\affiliation{Center for Astrophysics and Space Astronomy, University of Colorado, 389 UCB, Boulder, CO 80309, USA}

\received{January 31, 2018}
\revised{March 3, 2018}
\accepted{March 5, 2018}

\begin{abstract}

We measure \HH~temperatures and column densities across the Orion BN/KL explosive outflow from a set of thirteen near-IR \HH~rovibrational emission lines observed with the TripleSpec spectrograph on Apache Point Observatory's 3.5-meter telescope. We find that most of the region is well-characterized by a single temperature ($\sim$2000--2500 K), which may be influenced by the limited range of upper energy levels (6000--20,000 K) probed by our data set. The \HH~column density maps indicate that warm \HH~comprises 10$^{-5}$--10$^{-3}$ of the total \HH~column density near the center of the outflow. Combining column density measurements for co-spatial \HH~and CO at $T$ = 2500 K, we measure a CO/\HH~fractional abundance of 2$\times$10$^{-3}$, and discuss possible reasons why this value is in excess of the canonical 10$^{-4}$ value, including dust attenuation, incorrect assumptions on co-spatiality of the \HH~and CO emission, and chemical processing in an extreme environment. We model the radiative transfer of \HH~in this region with UV pumping models to look for signatures of \HH~fluorescence from \ion{H}{1}~\Lya~pumping. Dissociative (J-type) shocks and nebular emission from the foreground Orion \HII~region are considered as possible \Lya~sources. From our radiative transfer models, we predict that signatures of \Lya~pumping should be detectable in near-IR line ratios given a sufficiently strong source, but such a source is not present in the BN/KL outflow. The data are consistent with shocks as the \HH~heating source.

\end{abstract}
\section{Introduction} \label{sec:Introduction}

The Orion BN/KL outflow is a $\sim$500 year-old explosion of dense gas emanating from the Orion Molecular Cloud (OMC1) core, likely caused by the dynamical decay or merger of a multiple system of massive stars \citep{Bally2005,Bally2017,Luhman2017}. Over one hundred well-collimated jet-like structures have been observed in IR \HH~emission (termed ``\HH~fingers"; \citealt{Bally2015}), and more continue to be found with high-resolution near-IR spectroscopy that can velocity-resolve \HH~fingers that are spatially coincident along our line of sight \citep{Oh2016}. The 3-dimensional structure of the dense gas as traced by CO in low-$J$ \citep{Zapata2009,Bally2017}, mid-$J$ \citep{Peng2012a}, and high-$J$ \citep{Goicoechea2015} emission all indicate that CO appears to closely follow the inner \HH~fingers \citep{Bally2017}. The fingers' morphology and velocity vectors trace back to a common origin at approximately the same time, indicating that the outflow is due to a single explosive event that occurred $\sim$500 years ago. Kinematic studies of the embedded massive stars near the outflow origin have determined that three or four massive young stars were within several hundred AU of each other at the time of the explosion (although it is uncertain which stars were coincident), indicating a dynamical interaction as the impetus (e.g., \citealt{Tan2004,Gomez2008,Plambeck2009,Goddi2011,Chatterjee2012,Dzib2017,Plambeck2016,Rodriguez2017}).

Located just behind the Orion Nebula at a distance of 414$\pm$7 pc \citep{Menten2007}, most of the BN/KL outflow is heavily reddened with visual extinctions ranging from 3--10 mag \citep{Youngblood2016a}. The majority of the \HH~fingers associated with the BN/KL outflow can only be observed at longer wavelengths due to reddening. Some of the \HH~fingers (e.g., HH 201) are emerging from the dense OMC1 region into the foreground photo-dissociation region (PDR) and ionization front that separates the Orion blister \HII~region and OMC1, and they have been well studied at optical wavelengths \citep{Doi2004}. Because the \HH~fingers are bright, cover a large region of the sky ($\sim$3\arcmin~$\times$~3\arcmin), and have a curious origin, they have been extensively studied as a testbed for \HH~excitation mechanisms including shocks, UV fluorescence, and formation pumping (e.g., \citealt{Beckwith1978,Snell1984,Rosenthal2000,Bally2005,Colgan2007,Zapata2009,Bally2011,Bally2015,Youngblood2016a,Oh2016,Bally2017,Geballe2017}). The consensus is that no single shock model can satisfactorily explain the \HH~observations; there appears to be a superposition of post-shock temperatures and C- and J-type shocks with varying magnetic field strengths. Thorough observational constraints are needed to drive the advancement of shock models.

The \HH~fingers are also important to study because dynamical interactions between massive stars in young star forming regions are likely common, and the large amounts of energy released ($\sim$10$^{47}$ erg for BN/KL; \citealt{Kwan1976,Snell1984}) could be an important source of feedback in molecular clouds. Other possible BN/KL-like events include DR21 \citep{Zapata2013}, W49 Source G \citep{Smith2009}, G34.26+0.15 \citep{Cyganowski2008}, IRAS 05506+2414 \citep{Sahai2008}, and NGC 7129 \citep{Eisloffel2000,Gutermuth2004}, all in the Milky Way galaxy, and SPIRITS 14ajc in M83 \citep{Kasliwal2017}. However, these other systems are more distant ($\textgreater$~1 kpc), making BN/KL the ideal prototype for detailed study of this potentially common phenomenon.

Many previous studies have analyzed the excitation mechanisms of the \HH~fingers by comparing the flux ratios of multiple \HH~transitions (e.g., \citealt{Rosenthal2000,Colgan2007,Oh2016}), although no set of observations have yet obtained full spatial coverage of the outflow with high spatial resolution and high spectral resolution. These past studies argue that throughout the outflow, the \HH~emission is characterized by warm temperatures (2000--3000 K), consistent with shock excitation, with evidence of small contributions from fluorescence from stellar, accretion, or nebular emission. Recently, a hot temperature component ($T$~$\sim$~5000 K) comprising a few percent of the \HH~population has been observed in BN/KL and is likely due to \HH~formation pumping \citep{Geballe2017}. \HH~formation occurs predominantly on dust grains, as gas-phase formation is slow. Some of the 4.5 eV \HH~binding energy is kept by the \HH~molecule after it is ejected from the dust grain, thus ``pumping" it into excited ($v$,$J$) states and indicating temperatures larger than the molecular dissociation temperature ($T_{\rm dissoc}$~$\sim$~4500 K). The presence of formation pumping indicates that the shocks are at least somewhat dissociative, and thus should emit high-energy radiation that could create a fluorescent \HH~population.

Non-thermal \HH~populations due to radiative excitation from \Lya~photons (1215.67 \AA) have been observed in planetary nebulae \citep{Lupu2006}, accreting T Tauri stars \citep{Herczeg2002,Walter2003,France2012a}, reflection nebulae \citep{Le2017}, and several Herbig Haro objects \citep{Schwartz1983}. The source of \Lya~can either be nebular (\HI~recombination), from a central source (stellar or accretion), or intrinsic to the shock itself (\HI~recombination). In the recombination regions behind bow shocks, cooling is dominated by Lyman and Balmer \HI~transitions, producing strong \Lya~emission. \Lya~spectrally coincides with strong Lyman band \HH~transitions, but these transitions pump out of rotationally and/or vibrationally excited states ($E$~$\textgreater$~10,000 K; typically $v$~$\geq$~2). Thus the \HH~population must be hot ($T$~$\sim$~2000 K) to absorb \Lya~photons.

Using the near-IR position-position-velocity (PPV) cube from \cite{Youngblood2016a} with 1 square arcsecond spaxels and 86 km s$^{-1}$ spectral resolution that covers the entire BN/KL outflow (2.7\arcmin~$\times$~3.3\arcmin), we determine the temperatures and column densities on a spaxel-by-spaxel basis from 13 observed \HH~emission lines. We also look for signatures of radiative excitation from nebular \Lya~and intrinsic shock \Lya, which leave a non-thermal signature in the level populations of the ground electronic state.

Section~\ref{sec:ObservationsReductions} briefly describes the near-IR PPV cube and data products used for this analysis. Section~\ref{sec:Rovib_temps} describes the temperatures and column densities determined across the outflow, and Section~\ref{sec:CO} compares the \HH~observations to previous CO observations, including a CO/\HH~abundance measurement for the hot gas. Section~\ref{sec:nonthermal} discusses the signatures of non-thermal populations due to \Lya~pumping, and Section~\ref{sec:Summary} summarizes the results.

\section{Near-IR \HH~images} \label{sec:ObservationsReductions}

We utilize a near-IR (1.1--2.4~\mum) position-position-velocity (PPV) cube of the Orion BN/KL outflow (2.7\arcmin~$\times$~3.3\arcmin) that was assembled from $R$ =  3500 spectra from Apache Point Observatory's cross-dispersed TripleSpec spectrograph. The observations, data reduction, and the creation of derivative data products are described fully in \cite{Youngblood2016a} and are available for public use\footnote{The J, H, and K band PPV cubes are available for download at http://dx.doi.org/10.7910/DVN/YUNZ1F.}. The PPV cubes have a dispersion of 2.88 \AA~pix$^{-1}$ and a spatial scale of 1\arcsec~pix$^{-1}$. The derivative data products from the PPV cube include integrated intensity maps (observed and de-reddened) of 13 \HH~emission lines, a visual extinction map, radial velocity maps, and linewidth maps. 

We use the de-reddened integrated intensity maps of the 13 \HH~emission lines in this work (Table~\ref{table:H2_lines}). Integrated intensities were measured by fitting Gaussians to the observed line profiles in the PPV cube and reddening was corrected for using the 1-0 Q(3) (24237 \AA) and 1-0 S(1) (21218 \AA) emission lines that originate from the same upper ($v$,$J$) state. The 13 lines vary significantly in S/N, thus each line has varying spatial coverage across the mapped region. The inner regions of the outflow are the brightest (Peaks 1 and 2; \citealt{Beckwith1978}) and have the most spectral line coverage.

Deep \HH~observations by \cite{Pike2016} and \cite{Geballe2017} have revealed that the 3-2 S(3) line (22014 \AA) is significantly contaminated by the 4-3 S(5) line. The two lines are separated by 55 km s$^{-1}$, which is unresolved in the PPV cube (dispersion of 39 km s$^{-1}$ pix$^{-1}$ at 22014 \AA), and the 4-3 S(5) line contributes $\sim$25\% of the flux \citep{Pike2016}. We attempted to simultaneously fit one Gaussian for each line, but the fits were unsatisfactory. Thus, we have systematically decreased the flux in our 3-2 S(3) map by 25\%. 

For use in Section~\ref{sec:nebular_lya}, we create a second visual extinction map based on the flux ratio between \HI~Br$\gamma$~(21661 \AA) and \HI~Pa$\beta$~(12818 \AA). Under Case B recombination at $T$ = 10$^4$ K and $n_{\rm H}$ = 10$^{4}$ cm$^{-3}$, the intrinsic flux ratio $F$(Br$\gamma$)/$F$(Pa$\beta$) = 0.17 \citep{Hummer1987}. We assume the near-IR extinction law from \cite{Mathis1990}. We compared this extinction map with that created from VLT/MUSE observations of H$\alpha$~and H$\beta$~\citep{Weilbacher2015}, and find them to agree within $A_{\rm V}$~$\sim$~1 mag.

\begin{deluxetable}{cccccc}
\tablecolumns{6}
\tablewidth{0pt}
\tablecaption{ Observed \HH~emission lines  \label{table:H2_lines}} 
\tablehead{\colhead{Transition} & 
                  \colhead{($v_u$,$J_u$) $\rightarrow$~($v_l$,$J_l$)} &
                  \colhead{$E_{u}$ (K)} &
                  \colhead{$\lambda_0$ (\AA)} &
                  \colhead{$A_{ul}$} &
                  \colhead{\Lya~cascade$^{\dagger}$} \\
                  \colhead{} &
                  \colhead{} &
                  \colhead{} & 
                  \colhead{} & 
                  \colhead{(10$^{-7}$ s$^{-1}$)} &
                  \colhead{(\%)} 
                  }
\startdata
1-0 S(0) & (1,2)$\rightarrow$(0,0) & 6471 & 22235 & 2.53 & 4.0 \\
1-0 S(1) & (1,3)$\rightarrow$(0,1) & 6951 & 21218 & 3.47 & 11.6 \\
1-0 S(7) & (1,9)$\rightarrow$(0,7) & 12823 & 17480 & 2.98 &  0.08 \\
1-0 S(8) & (1,10)$\rightarrow$(0,8) & 14233 & 17147 & 2.34 & 0.40 \\
1-0 S(9) & (1,11)$\rightarrow$(0,9) & 15747 & 16877 & 1.68 & 0.001 \\
2-1 S(0) & (2,2)$\rightarrow$(1,0) & 12095 & 23556 & 3.68 & 1.8 \\
2-1 S(1) & (2,3)$\rightarrow$(1,1) & 12550 & 22477 & 4.98 & 7.4 \\
3-2 S(3) & (3,5)$\rightarrow$(2,3) & 19086 & 22014 & 5.65 & 3.3 \\
3-2 S(5) & (3,7)$\rightarrow$(2,5) & 20858 & 20656 & 4.53 & 0.27 \\
1-0 Q(1) & (1,1)$\rightarrow$(0,1) & 6149 & 24066 & 4.29 & 9.1 \\
1-0 Q(2) & (1,2)$\rightarrow$(0,2) & 6471 & 24134 & 3.03 & 4.7 \\
1-0 Q(3) & (1,3)$\rightarrow$(0,3) & 6951 & 24237 & 2.78 & 9.3 \\
1-0 Q(4) & (1,4)$\rightarrow$(0,4) & 7584 & 24375 & 2.65 & 6.3 \\
\enddata
\tablecomments{Upper level energies ($E_{u}$) and wavelengths ($\lambda_0$) calculated via \citealt{Herzberg1950}. Einstein A coefficients ($A_{ul}$) are from \citealt{Wolniewicz1998}.}
\tablenotetext{\dagger}{The \Lya~cascade column denotes what percentage of the flux pumped from the (2,5) and (2,6) levels cascades through these lines. The ortho transitions' percentages are calculated by normalizing to the flux pumped out of (2,5) and the para transitions' percentages are calculated by normalizing to the flux pumped out of the (2,6) level of the ground electronic state. See Table~\ref{table:otherlines} for a complete list of strong \HH~lines and the percentage of the \Lya~cascade flux that appears in each line.}
\end{deluxetable}

\section{rovibrational Temperatures and Column Densities} \label{sec:Rovib_temps}

We determine the \HH~temperatures on a spaxel-by-spaxel basis across the BN/KL outflow. Using the de-reddened intensities $I$($v_u$,$J_u$ $\rightarrow$ $v_l$, $J_l$) in erg cm$^{-2}$ s$^{-1}$ sr$^{-1}$ of our 13 \HH~transitions, we calculate the \HH~column density in the upper states as

\begin{equation}
N(v_u,J_u) = \frac{4\pi \lambda_0}{hc} \frac{I(v_u,J_u \rightarrow v_l, J_l)}{A(v_u,J_u \rightarrow v_l, J_l)},
\label{eq:N_upper}
\end{equation}

\noindent where $A$($v_u$, $J_u$ $\rightarrow$ $v_l$, $J_l$) is the Einstein-A coefficient from \cite{Wolniewicz1998}. Equation~\ref{eq:N_upper} assumes that the \HH~emission is optically thin. Because the transition probabilities of the quadrupole transitions are small ($\sim$10$^{-7}$ s$^{-1}$; Table~\ref{table:H2_lines}), the optical depth remains much less than unity until \HH~column densities approach 10$^{24}$ cm$^{-2}$, which is orders of magnitude larger than the level column densities observed by previous authors for the BN/KL outflow, such as $N$($v$,$J$)~$\lesssim$~10$^{18}$ cm$^{-2}$ from \cite{Rosenthal2000}. However, we note that the measured column densities from \cite{Rosenthal2000} using the \textit{Infrared Space Observatory} are potentially lower limits due to beam dilution. 

We divide the level column densities $N$($v_u$,$J_u$) by their level degeneracy $g_{J_u}$ = $g_s$(2$J_u$+1), where $g_s$ = 3 for ortho (odd $J$) \HH~and $g_s$ = 1 for para (even $J$) \HH, and construct excitation diagrams by plotting the logarithm of $N$($v_u$,$J_u$)/$g_{J_u}$ against the upper level energy $E$($v_u$,$J_u$)/$k_B$ (K). The upper energy levels are calculated using equations and constants from \cite{Herzberg1950}. Temperatures and ($v$,$J$) = (0,0) column densities are derived from linear fits to the excitation diagrams, which is based on the Boltzmann equation:

\begin{equation}
\log_{10} \frac{N(v_u,J_u)}{g_{J_u}} = -\frac{1}{T \cdot \ln 10} \times \frac{E(v_u,J_u)}{k_B} + \log_{10} N(0,0).
\label{eq:H2_temp_log10}
\end{equation}

\begin{figure}
   \begin{center}
   
     \subfigure{
          \includegraphics[width=\textwidth]{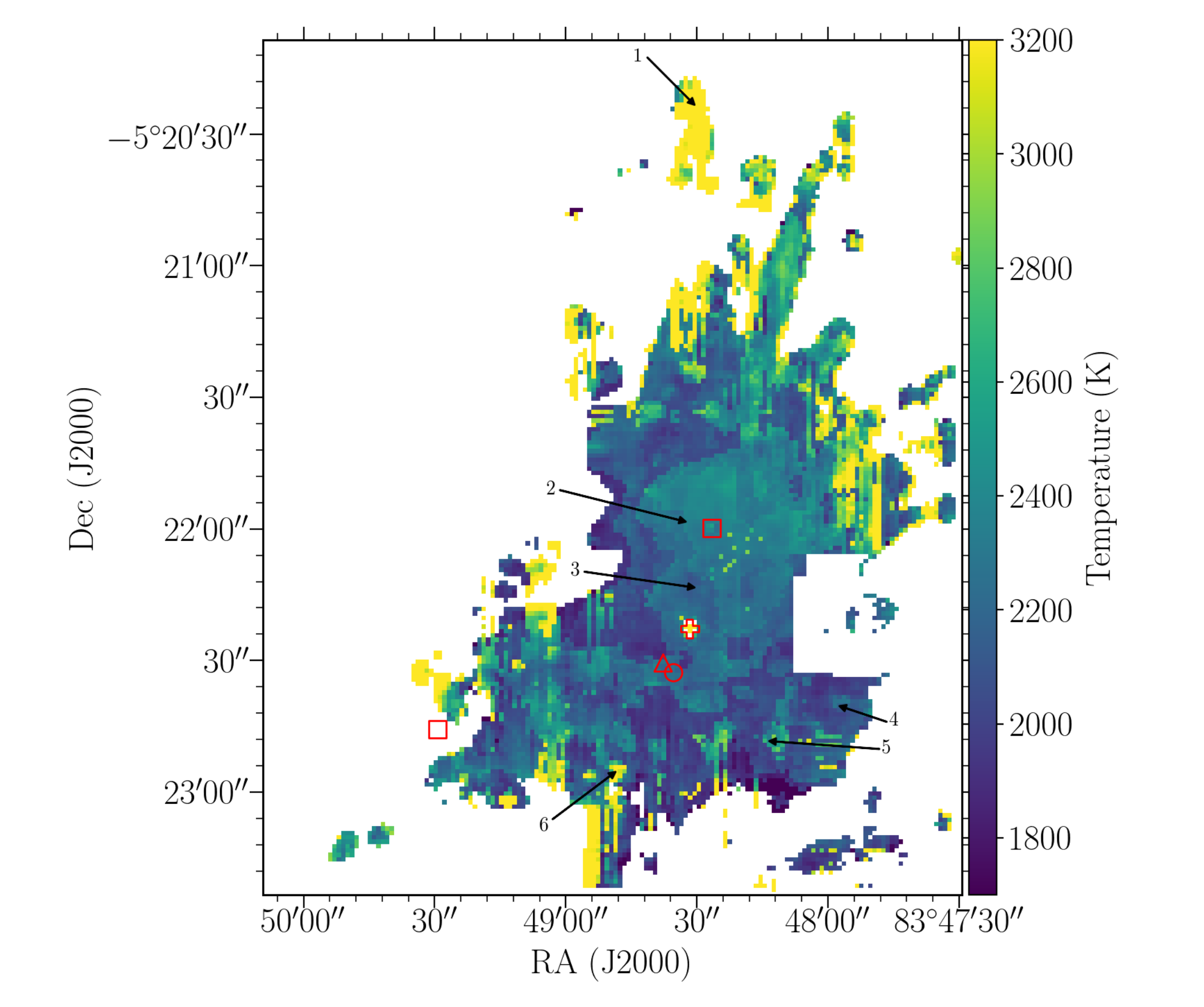}
          }

   \end{center}
    \caption{\HH~temperatures derived from the de-reddened \HH~intensities of all the available lines. The red plus sign, triangle, and open circle mark the locations of the BN object, source I, and source n, respectively. The red squares mark the locations of V2248 Ori (northwest) and MT Ori (southeast), continuum sources that contaminate some of the \HH~emission line maps. The labeled arrows show the spaxel locations of the sample fits shown in Figure~\ref{fig:example_h2_fits}. The data behind this figure is available in the online journal.
        }
    \label{fig:H2_temp}

\end{figure}  

\begin{figure}
   \begin{center}
   
     \subfigure{
          \includegraphics[width=\textwidth]{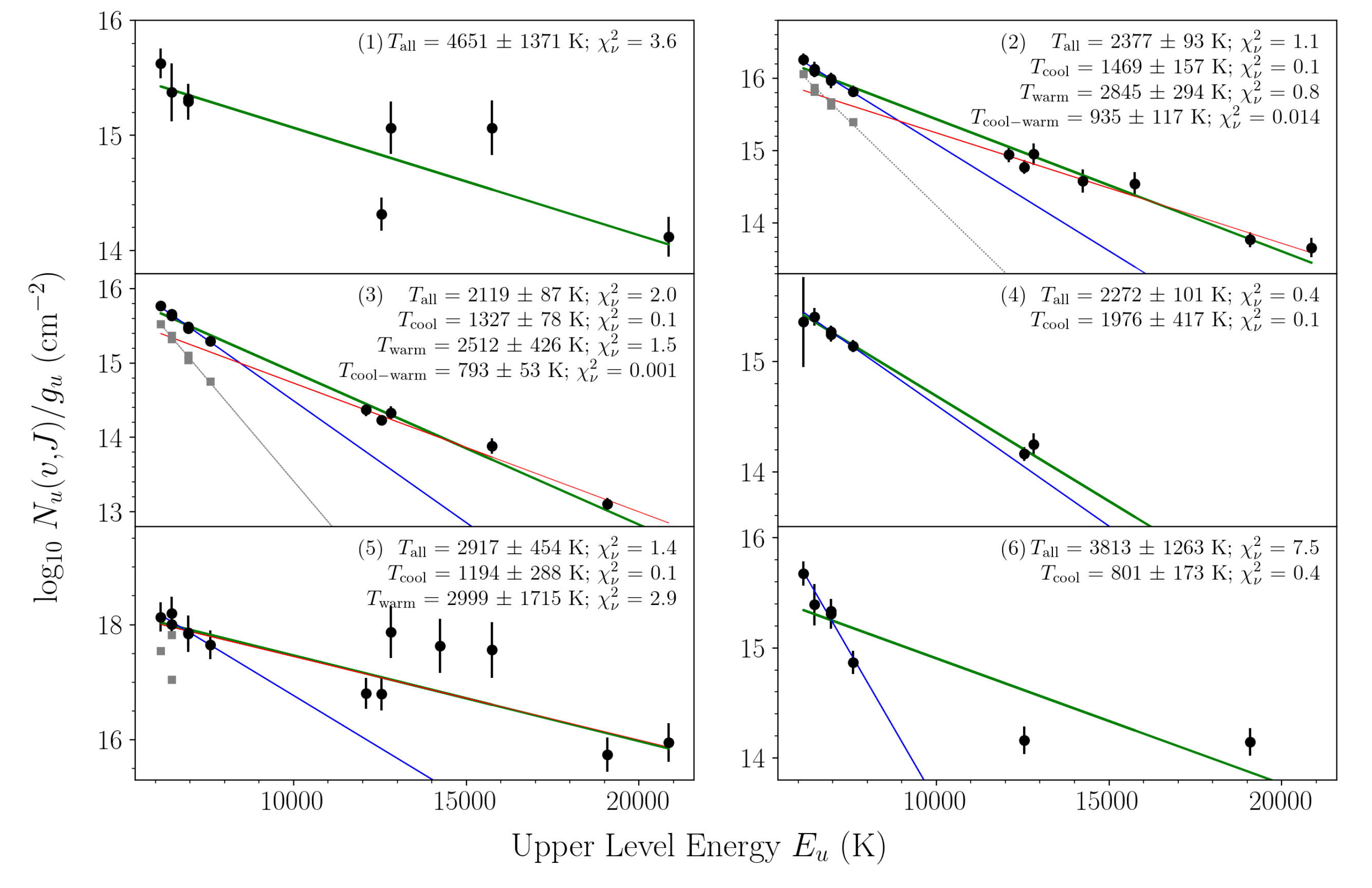}
          }

   \end{center}
    \caption{Representative \HH~temperature fits from six different 1\arcsec~$\times$~1\arcsec~spaxels of the PPV cube. The numbers in parentheses correspond to the marked spaxels in Figure~\ref{fig:H2_temp} and subsequent figures. The black points with error bars show the data, the green line shows the fits to all available \HH~lines (``all"), the blue line shows the fits to lines with $E_u$~$\textless$~8,000 K (``cool"), and the red line shows the fits to lines with $E_u$~$\textgreater$~12,000 K (``warm"). The grey squares show the data for the ``cool" upper levels after the ``warm" component's contribution to those levels has been subtracted (``cool-warm"), and the grey dotted line shows the fit to the squares. Their error bars are large due to the uncertainty in the subtracted ``warm" fit and are not shown for visual clarity. The best-fit temperature, uncertainty, and reduced chi-square value for each case is printed at the top of each diagram. 
        }
    \label{fig:example_h2_fits}

\end{figure}

\begin{figure}
   \begin{center}
   
     \subfigure{
          \includegraphics[width=\textwidth]{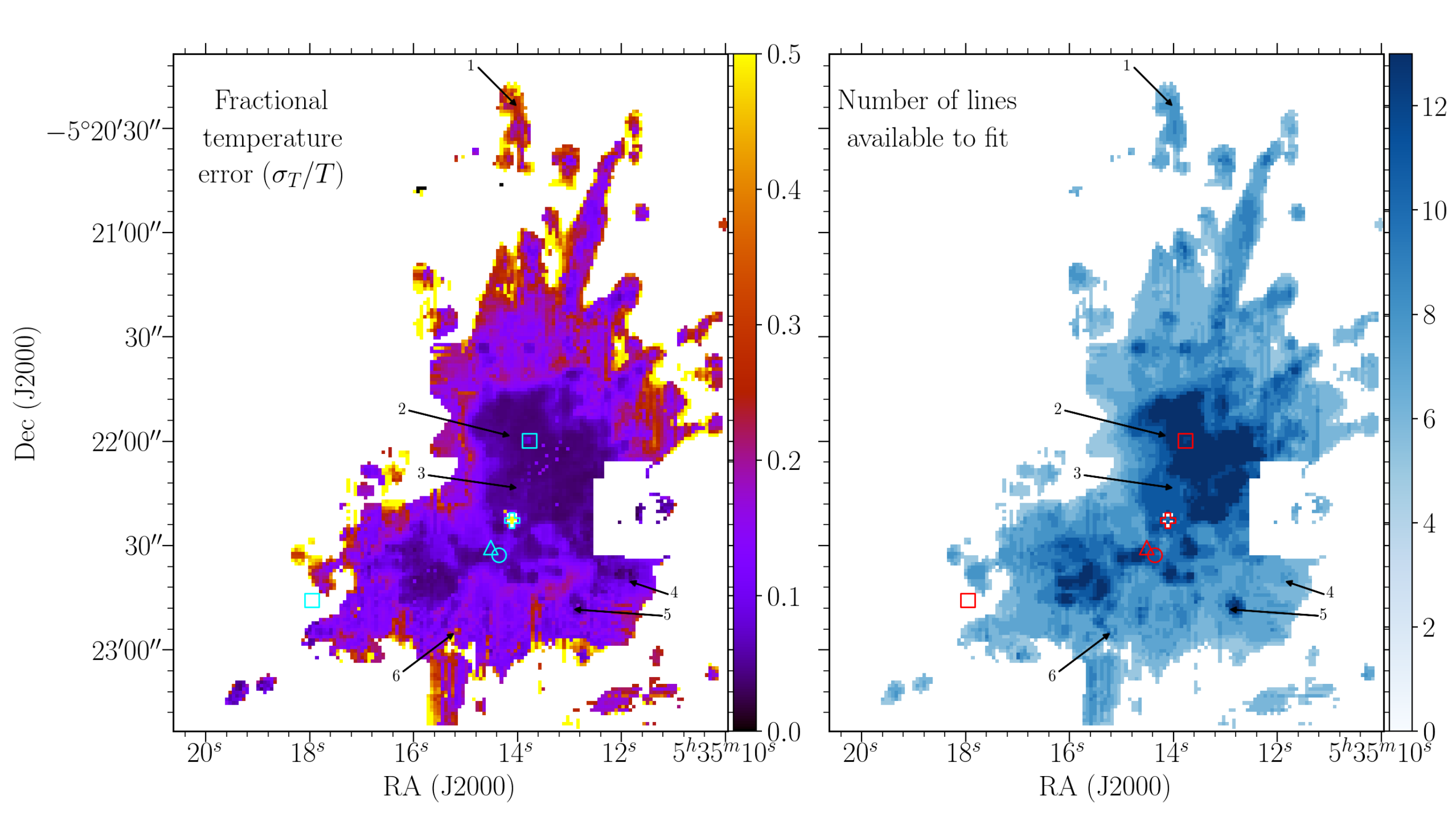}
          }

   \end{center}
    \caption{\textit{Left:} The fractional uncertainty in the \HH~temperature, $\sigma_T$ divided by the temperature $T$. \textit{Right:} The number of \HH~emission lines used in the temperature fit for each spaxel. The minimum is 3 and the maximum is 13.
        }
    \label{fig:h2_all_temp_eval_plot}

\end{figure}

\noindent Total column density is recovered from $N$(0,0) by assuming a thermal distribution at the fitted temperature $T$. To compute a temperature via a linear fit for a given spaxel, we require at least three \HH~line fluxes from transitions originating from different upper energy levels. Of the 21,270 1\arcsec$~\times$~1\arcsec~spaxels covered by the PPV cube, 11,246 spaxels contain at least three \HH~line fluxes, but only 9,302 spaxels contain three \HH~line fluxes from transitions that do not share an upper state. 

In Figure~\ref{fig:H2_temp}, we present a temperature map based on fitting one temperature component, where we find that most temperatures are between 2000--3000 K. Figure~\ref{fig:example_h2_fits} shows example temperature fits for marked spaxels in Figure~\ref{fig:H2_temp}, and Figure~\ref{fig:h2_all_temp_eval_plot} shows the fractional temperature errors ($\sigma_{\rm T}$/$T$) and the number of \HH~lines fitted. In the bright, central regions of the outflow (Peaks 1 and 2), the temperature errors are at their smallest: 3--5\%. Most of the temperatures have uncertainties $\textless$20\%, and larger uncertainties typically correspond to fits with fewer available \HH~lines. 

When using the observed \HH~intensities, rather than de-reddened values, calculated temperatures are underestimated. The higher $E_{\rm u}$~lines in the dataset are at shorter wavelengths and are therefore more reddening sensitive. In the highest S/N regions of the PPV cube, the temperatures are underestimated by 100--400 K, which is significant when compared to the absolute temperature uncertainties in that region. Elsewhere, the temperature uncertainty is comparable to the difference between temperatures derived from observed and de-reddened intensities, meaning the reddening correction is not critically important in low-S/N data where other uncertainties dominate.

Some of the temperature fits in Figure~\ref{fig:example_h2_fits} indicate the presence of \HH~populations that could be characterized by two temperatures. In many instances, the $J$ $\leq$~4 lines in the $v$ = 1 state (6000 K~$\textless$~$E_u$~$\textless$~8000 K) indicate cooler temperatures (steeper slopes) than found by the fits that include $E_u$~$\textgreater$~12,000 K lines. We create two additional temperature maps: one based on the six emission lines originating from the 6000 K~$\textless$~$E_u$ $\textless$~8000 K states, and one based on the seven emission lines originating from the 12,000 K~$\textless$~$E_u$ $\textless$~21,000 K states. The warm component contributes significantly to the cool component (Figure~\ref{fig:example_h2_fits}), so we extrapolate the warmer lines' fit to the upper energy levels of the cooler lines and subtract the warm component. This leads to lower temperatures and higher column densities for the cool component than would be found without correcting for the warm component. 

\begin{figure}
   \begin{center}
   
     \subfigure{
          \includegraphics[width=\textwidth]{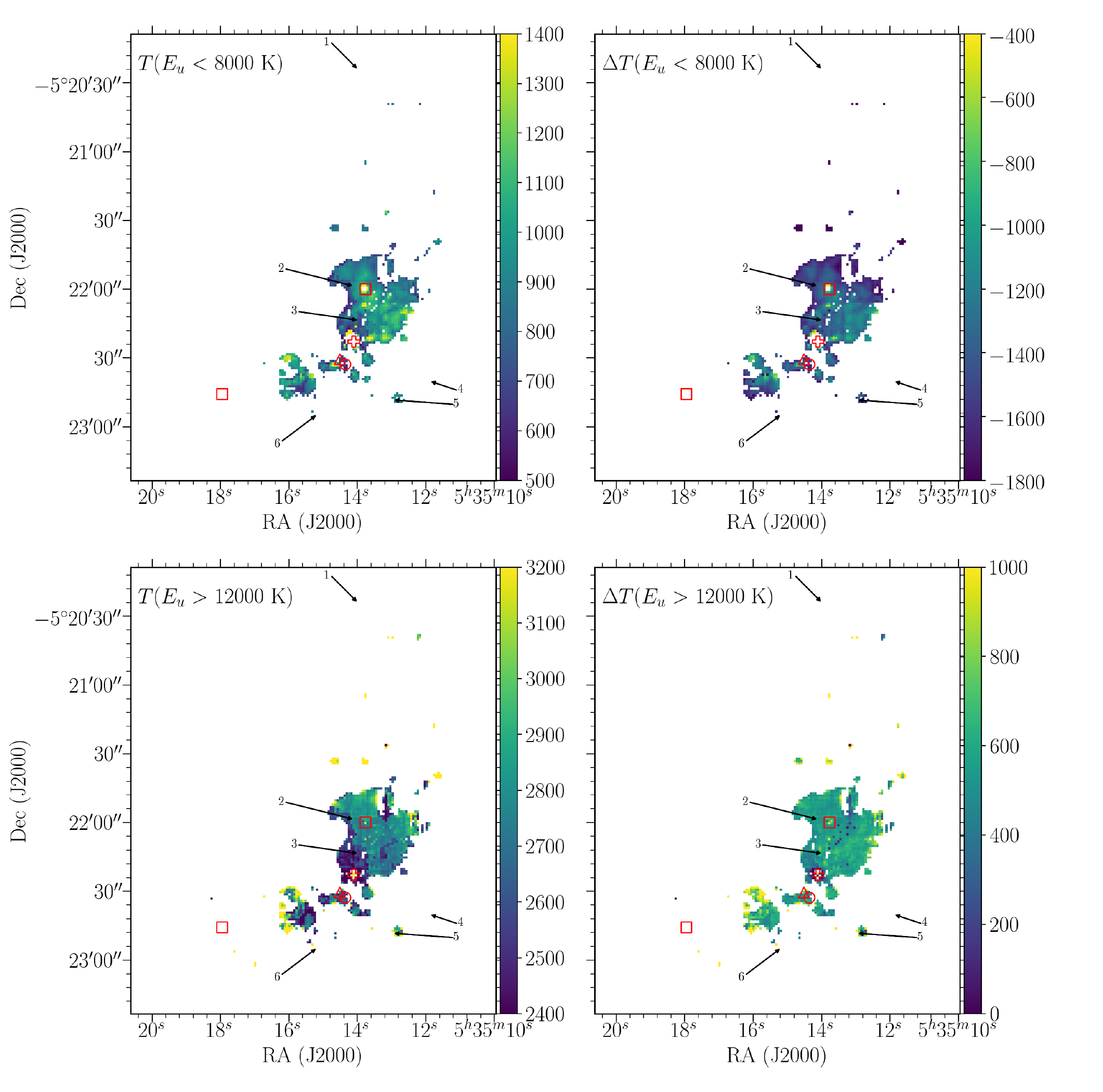}
          }

   \end{center}
    \caption{\textit{Top Left:} The temperature (K) derived using only the \HH~emission lines that originate from upper levels with $E_u$~$\textless$~8000 K after the fit to the $E_u$~$\textgreater$~12,000 K lines has been subtracted. \textit{Top Right:} The cooler $E_u$~$\textless$~8000 K temperatures with the hotter ``average" temperature (derived using all of the observed lines; Figure~\ref{fig:H2_temp}) subtracted. The colorbar shows the difference in the two temperature maps in Kelvin. \textit{Bottom Left:} The temperature (K) derived using only the $E_u$~$\textgreater$~12,000 K lines. \textit{Bottom Right:} The warmer $E_u$~$\textgreater$~12,000 K temperatures with the cooler ``average" temperature subtracted.
        }
    \label{fig:h2_temperature_4panel_comparison}

\end{figure}

Using the cooler lines, we find significantly cooler temperatures across the outflow ($\sim$1400 K; Figure~\ref{fig:h2_temperature_4panel_comparison}). The morphology of this cooler temperature map is significantly different than the morphology of the temperature map in Figure~\ref{fig:H2_temp} that uses lines from all of the states; it traces the 1-0 S(1) emission well. Because the cooler temperature component traces the \HH~fingers well, its excitation should be due to the shocks. Using the warmer lines, we find that the temperatures are on average 500 K warmer than the temperatures found using all the available lines. The morphology of this temperature map more closely resembles the central, smooth temperature distribution of Figure~\ref{fig:H2_temp}, which is based off of all lines.

We consider but discard the possibility that this morphological smoothness of the warmer regions is due to scattering of the \HH~emission from regions heated by the region's embedded massive protostars. \cite{Burton1991} found a \HH~reflection nebula spatially coincident with Peaks 1 and 2, but concluded that the observed dichroic polarization is due to absorption from a foreground layer of aligned dust grains. \cite{Sugai1994} confirm that the polarized \HH~emission seen in Peaks 1 and 2 is intrinsic to the area and is polarized by foreground dust grains as opposed to being scattered into the line of sight.

\begin{figure}
   \begin{center}
   
     \subfigure{
          \includegraphics[width=\textwidth]{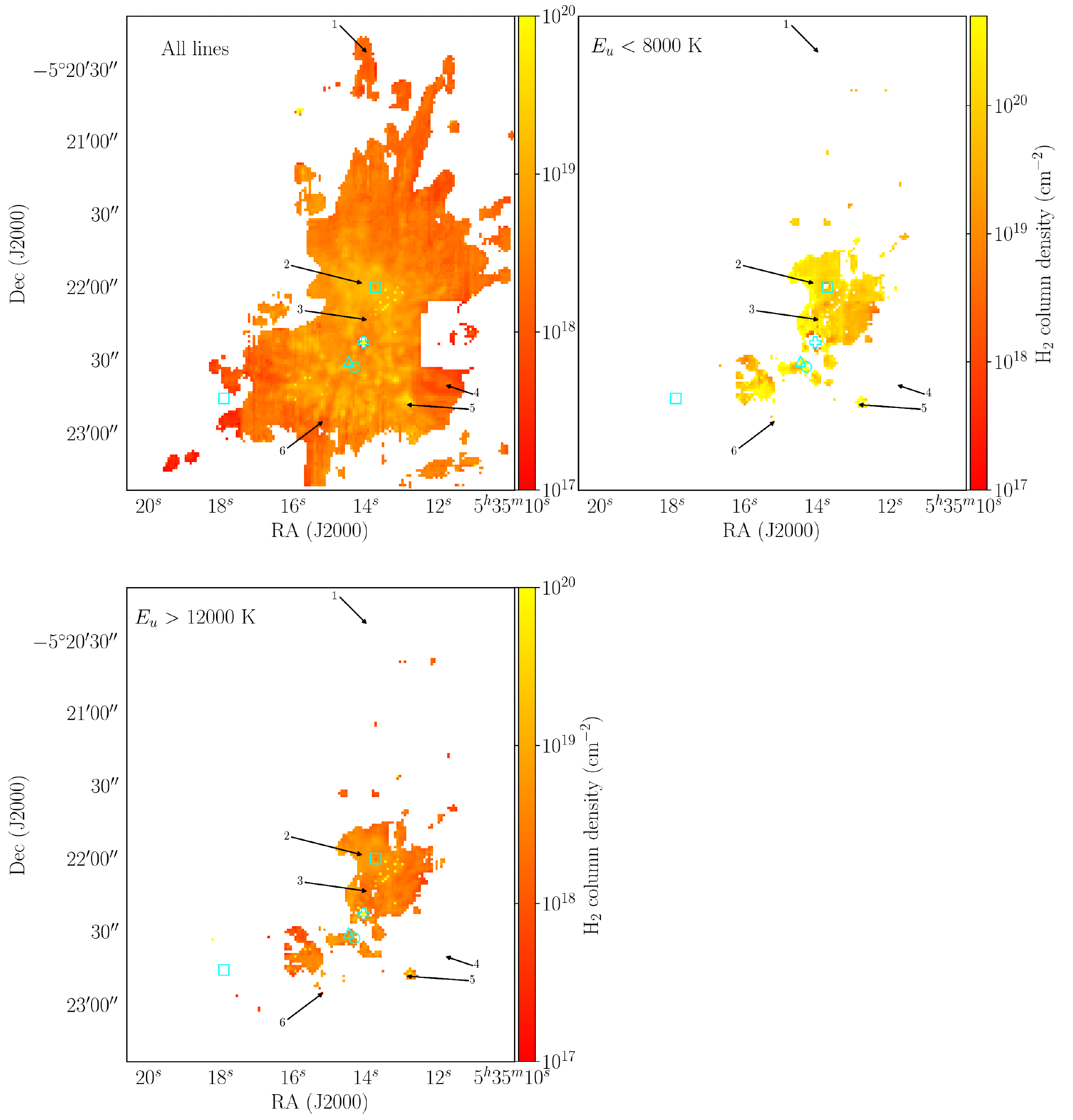}
          }

   \end{center}
    \caption{Total \HH~column density maps that correspond to the three temperature maps. Each spaxel's value represents the column density of \HH~at that spaxel's temperature. \textit{Top left:} Derived from fits to all 13 \HH~lines. Data behind this figure is available in the online journal. \textit{Top right:} Derived from fits to the $E_u$~$\textless$~8000 K lines after the fit to the $E_u$~$\textgreater$~12,000 K lines has been subtracted. \textit{Bottom left:} Derived from fits to only the $E_u$~$\textgreater$~12,000 K lines.
        }
    \label{fig:h2_column_3panel_comparison}

\end{figure}

From the excitation diagrams, we also measure the total column density of the \HH~population at the best-fit temperatures, which we refer to as $N$(\HH, hot) to distinguish it from the total column density of \HH~at all temperatures (Figure~\ref{fig:h2_column_3panel_comparison}). For the fits using all 13 \HH~lines, $N$(\HH, hot) values range from approximately 10$^{17}$ -- 10$^{19}$ cm$^{-2}$, and larger columns trace the \HH~finger structure towards the center of the outflow. Larger $N$(\HH, hot) values are found when fitting only the $E_u$~$\textless$~8000 K lines, and these values range from 10$^{19}$ -- 10$^{21}$. This indicates that cooler gas dominates the \HH~population. The $N$(\HH, hot) values found when fitting only the $E_u$~$\textgreater$~12,000 K lines are only slightly smaller than the values found from fitting all 13 lines. Using the dust extinction map from \citep{Youngblood2016a} and the dust-to-gas ratio from \citep{Diplas1994}, we measure the total foreground \HH~column density at all temperatures to be $\sim$10$^{22}$ cm$^{-2}$. Therefore, the rovibrationally excited or hot \HH~comprises 10$^{-5}$--10$^{-3}$ of the total foreground \HH~column density.

\subsection{Potential biases and systematic errors} \label{sec:biases}

\begin{figure}
   \begin{center}

               \subfigure{
          \includegraphics[width=\textwidth]{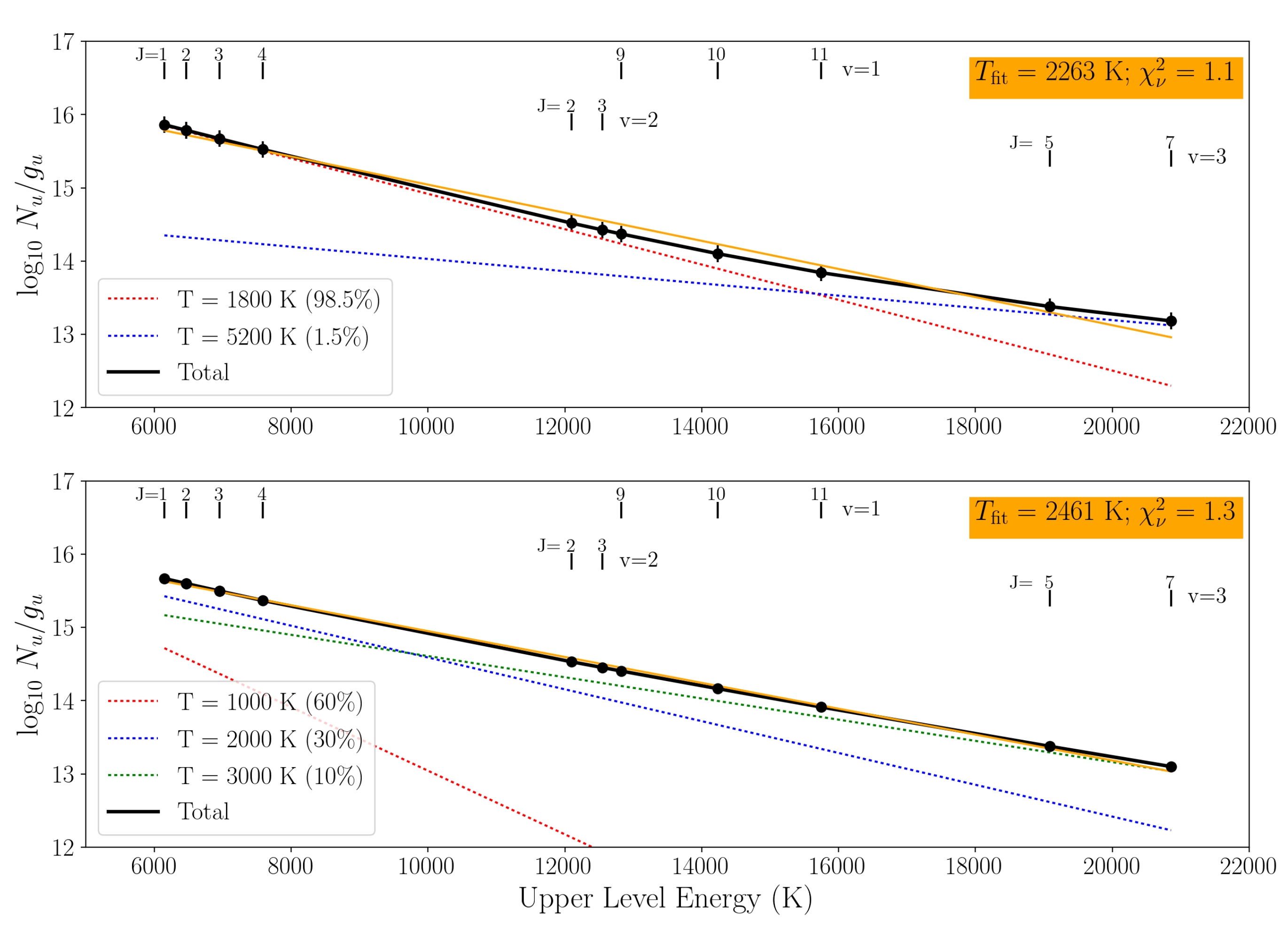}
          }

   \end{center}
    \caption{Example excitation diagrams of multi-temperature \HH~populations using our 13 near-IR \HH~lines. For both panels, the total \HH~column density was set to log$_{10}$ $N$(\HH) = 19 and ortho/para = 3. \textit{Top:} 98.5\% of the \HH~population is described by $T$ = 1800 K (red dashed line), and 1.5\% is described by $T$ = 5200 K (blue dashed line) as \cite{Geballe2017} found for part of the BN/KL outflow. The orange line shows a single temperature fit to the 13 points: $T$ = 2263 K and $N$(\HH) = 5.5$\times$10$^{18}$ cm$^{-2}$ with \redchi~= 1.1, indicating a good fit. \textit{Bottom:} 60\% of the \HH~population has $T$ = 1000 K, 30\% has $T$ = 2000 K, and 10\% has $T$ = 3000 K. The orange line shows a single temperature fit to the 13 points: $T$ = 2461 K and $N$(\HH) = 3.5$\times$10$^{18}$ cm$^{-2}$ with \redchi~= 1.3, indicating a good fit.
        }
    \label{fig:fake_excitation_diagrams}

\end{figure}

The set of 13 \HH~lines do not sample an upper energy level range suitable to detect multiple temperature components. In Figure~\ref{fig:fake_excitation_diagrams}, we show two examples of realistic multi-temperature thermal populations, where a single temperature fits well, as indicated by the reduced chi-square values. \cite{Pike2016} also noted for K band observations the necessity of a two-temperature fit is not obvious unless lines from levels $E_u$~$\geq$~30,000 K are included, or lines from levels $E_u$~$\textless$~6000 K \citep{Rosenthal2000}. 

\begin{figure}
   \begin{center}
   
     \subfigure{
          \includegraphics[width=\textwidth]{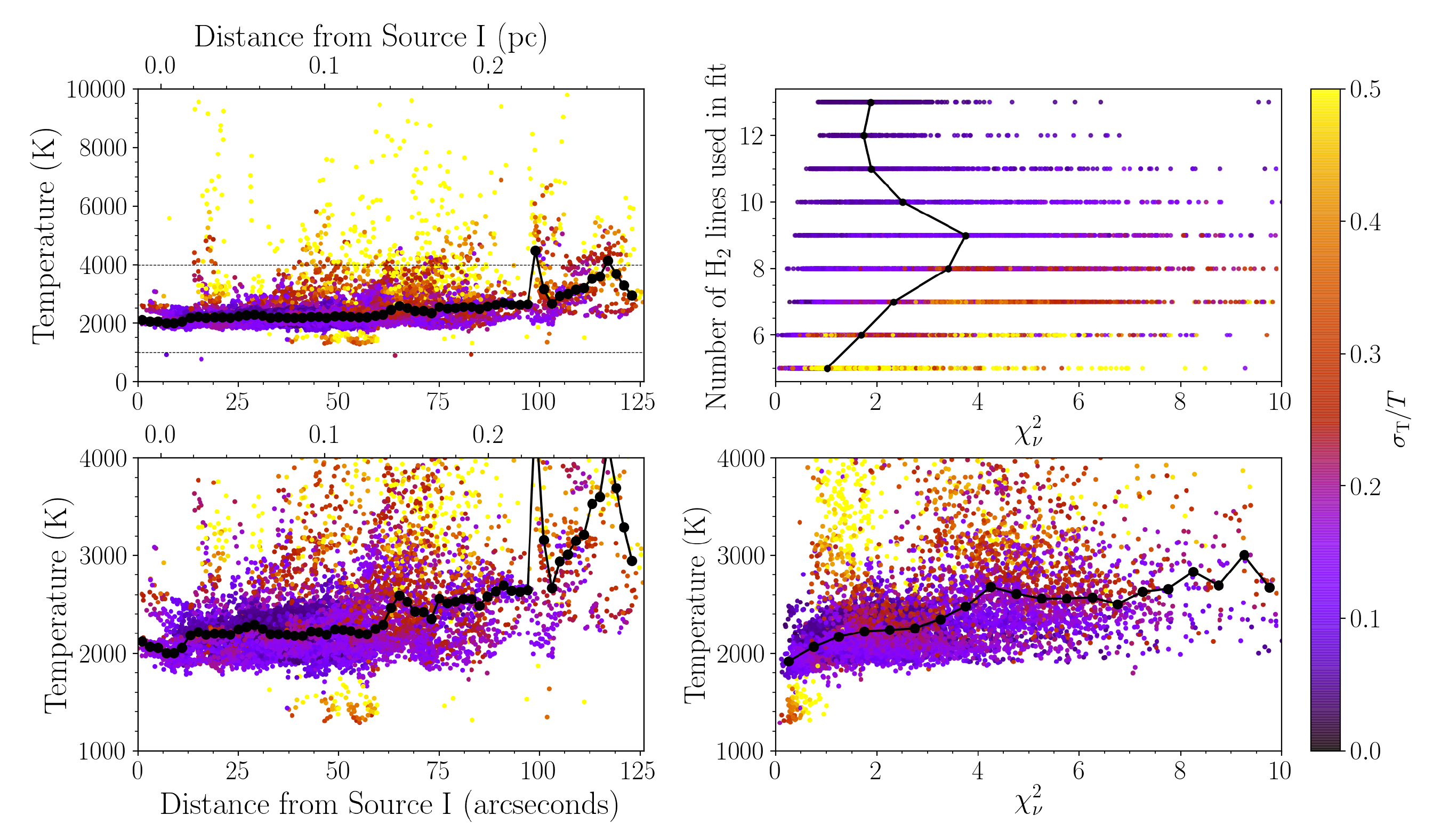}
          }

   \end{center}
    \caption{\textit{Top left:} Fitted temperature compared to the distance from Source I in arcseconds (bottom axis) and pc (top axis) assuming a distance of 414 pc. The dashed lines shows the temperature range corresponding to the bottom left plot. Each circle represents a spaxel and is color-coded by the fractional uncertainty in the fitted temperature ($\sigma_{\rm T}$/$T$). The black points represent the median values in each distance bin. \textit{Bottom left:} Similar to the top left plot, but shows a narrower temperature range as indicated by the dashed lines in the top left plot. \textit{Top right:} The reduced chi-square (\redchi) of the fit is shown against the number of \HH~lines used in the fit. \textit{Bottom right:} The fitted temperature is shown against \redchi.
        }
    \label{fig:diagnostic_plots}

\end{figure}

The fitted temperatures are likely biased by which of the 13 lines are available to fit and the S/N of each of those lines. Figure~\ref{fig:diagnostic_plots} demonstrates that the goodness-of-fit metric \redchi~is correlated with the fitted temperature (Pearson correlation coefficient $\rho$~= 0.91 and the probability of no correlation $n$ = 3$\times$10$^{-8}$). The median fitted temperatures rise by approximately 100 K between fits using five \HH~lines and fits using all thirteen \HH~lines. Also, temperature fits with small numbers of \HH~lines ($\leq$7) and large numbers of \HH~lines ($\geq$11) have the best fits as defined by the \redchi~value. This is likely because of our two distinct populations of \HH~lines. Six of our lines originate from $E_u$~$\textless$~8000 K and seven from $E_u$~$\textgreater$~12,000 K, leaving a wide gap in probed energy levels, and the latter lines are lower S/N. It is apparent that fitting more lines reduces the \redchi~value. Fitting seven or fewer lines also reduces the \redchi~value, because for spaxels with seven or fewer lines measured, the available lines tend to come from the $E_u$~$\textless$~8000 K population, which has higher S/N. It is possible that the fits to only the $E_u$~$\textless$~8000 K lines are more consistent with a single temperature than the fits that included the higher $E_u$ lines as well, but as shown in Figure~\ref{fig:fake_excitation_diagrams}, this is likely not the case.

We also observe an increase in temperature with increasing distance from the center of the outflow, which coincides with the position of Source I within a few arcseconds (Figure~\ref{fig:diagnostic_plots}). This temperature increase could be caused by faster shocks and/or lower volume density. Assuming a momentum-conserving explosion into a stationary medium, individual knots of ejected material become sorted by mass; the less massive knots attain higher velocities and travel farther from the origin. Higher outflow velocities will shock the gas to higher temperatures, potentially dissociating \HH~and leading to formation pumping and fluorescent excitation from radiative shocks. Also, the volume density $n$ of the gas is expected to be less in the outer fingers according to the momentum-conserving explosion scenario. This implies that the cooling time of the gas, which is inversely proportional to $n^2$, is longer in the outer fingers and could explain the higher temperatures. 

Thus, our relative temperature measurements are most robust in the central region of the outflow where there are more high-S/N lines available for fitting. Our absolute fitted temperatures are influenced by the available selection of \HH~lines and their upper level energies (6000 $\textless$~$E_u$~$\textless$~21,000), and they typically have larger errors in the outer regions of the outflow.

\section{Comparison to CO observations} \label{sec:CO}

We compare our \HH~temperature ($T$(\HH) $\sim$~2000--3000 K) and column density measurements ($N$(\HH) $\sim$~10$^{17}$--10$^{19}$ cm$^{-2}$) to ALMA and \textit{Herschel} CO observations from the literature. \cite{Bally2017} mapped the entire BN/KL outflow in $^{12}$CO $J$ = 2--1 in high spatial (1--2\arcsec) and velocity resolution (1.3 \kms). This emission traces cold CO gas ($T_{\rm cold}$ = 20--90 K) at column densities similar to \HH, but the CO column densities are underestimated due to spatial filtering of the interferometer. ALMA confirmed Submillimeter Array observations that showed that these CO ``streamers"  are highly collimated and spatially coincident with the \HH~fingers, but only out to $\sim$1\arcmin~\citep{Zapata2009}. Beyond $\sim$1\arcmin, the \HH~fingers exhibit large velocities ($v$~$\textgreater$~100 \kms) where shocks could completely dissociate the CO molecules. Alternatively, the dense clumps needed to drive these high-velocity shocks \citep{Bally2015} are not as massive as the inner clumps and could be below the detection threshold ($T_{\rm brightness}$~$\textless$~0.1 K). A momentum-conserving explosion like BN/KL will ensure that the lowest mass ejecta have the highest velocities and are therefore farthest from the explosion origin. 

On the other hand, \cite{Goicoechea2015} mapped the inner 2\arcmin~$\times$~2\arcmin~BN/KL outflow in $^{12}$CO up to $J$ = 48--47 at 12\arcsec~spatial resolution with \textit{Herschel}. They found that multiple temperature components were required to explain the observations, with the hottest temperature component having a similar kinetic temperature to the near-IR \HH~gas ($T_{\rm hot}$ = 2500 K). The near-IR \HH~and the hot CO observed by \textit{Herschel} have a similar temperature and morphology, making it likely that they are co-spatial parcels of gas \citep{Goicoechea2015}. 

We measure the CO/\HH~ratio of Peak 1 (the northwestern peak of \HH~intensity as seen at low spatial-resolution; \citealt{Beckwith1978}) from the \HH~column densities measured in this work and the CO column density from \textit{Herschel}. For the 2500 K CO component, \cite{Goicoechea2015} measured $N$(CO, hot) = 1.5$\times$10$^{16}$ cm$^{-2}$ over a 30\arcsec$\times$30\arcsec~box centered at $\alpha_{2000}$: 5$^{\rm h}$35$^{\rm m}$13$^{\rm s}$.6, $\delta_{2000}$: -5$^{\circ}$22\arcmin07\arcsec.9. For this same region, we measure a mean value of $N$(\HH, hot) = 7.3$\times$10$^{18}$ cm$^{-2}$ after excluding the BN object (a known continuum contaminant in our emission line maps; \citealt{Youngblood2016a}) and convolving the column density map derived from all \HH~lines to match the \textit{Herschel} spatial resolution. This results in CO/\HH~= 2$\times$10$^{-3}$, and we estimate the uncertainty to be a factor of two. 

Our measured CO/\HH~= 2$\times$10$^{-3}$ is an order of magnitude larger than those reported for cold, dense clouds (2$\times$10$^{-4}$; \citealt{Dickman1978,Lacy1994,Lacy2017}) and even larger than the value for diffuse and translucent clouds \citep{Burgh2007}. Previous authors have made CO/\HH~measurements of Peak 1 and the surrounding area, but for colder gas. \cite{Wilson1986} found 5$\times$10$^{-5}$ for $\sim$100 K gas, and \cite{Watson1985} found 1.2$\times$10$^{-4}$ for 750 K gas. Models are in agreement with 10$^{-7}$~$\lesssim$~CO/\HH~$\lesssim$~10$^{-4}$ \citep{vanDishoeck1988,Visser2009}, but the majority of the temperature parameter space probed by models is considerably colder than 2500 K. In models of protoplanetary disks in transition, \cite{Bruderer2013} finds some regions of parameter space (e.g., disks with small gas fractions) where CO/\HH~$\textgreater$~10$^{-3}$.

We consider the possibility that dust is hiding significant amounts of hot \HH~and thus raising the CO/\HH~ratio. \cite{Goicoechea2015} measured CO column density from $\lambda$~$\sim$~100 \mum, while this work is centered around $\lambda$~= 2 \mum. Assuming a standard extinction law from \cite{Mathis1990} and that CO is unaffected by dust extinction, we find that $A_{\rm V}$ = 17 mag would be required to increase the typical CO/\HH~= 2.7$\times$10$^{-4}$ \citep{Lacy1994} to 2$\times$10$^{-3}$ (this work). Using the \cite{Diplas1994} relation between $A_{\rm V}$ and $N$(H) and assuming $N$(H) $\sim$~2$N$(\HH), $A_{\rm V}$ = 17 mag corresponds to $N$(\HH) = 7.5$\times$10$^{21}$ cm$^{-2}$. Assuming a gas volume density of $n$(\HH) = 10$^{6}$ cm$^{-3}$ \citep{Bally2017} and spherical symmetry, a cloud of diameter 0.003 pc or 500 AU (1.5\arcsec~in projection at 414 pc) would result in $A_{\rm V}$~=17 mag. In this scenario, $A_{\rm V}$~= 17 mag of extincting material lies behind the observed hot \HH~with $A_{\rm V}$~$\sim$~5 mag in front of the observed \HH. This implies that about 90\% of the total hot \HH~and CO is behind the observed \HH~fingers. Other possible influences on our measured CO/\HH~ratio could be confused spatial stratification resulting from the large apertures of our measurements (e.g., \citealt{France2014}), and uncertainties in the CO column density derived from the \textit{Herschel} data. \cite{Bruderer2012} discuss how uncertainty in \HH~formation rates at high gas temperatures can affect modeled high-$J$ CO line intensities by a factor of a few, and therefore influence the derived column densities.

\section{Non-thermal populations} \label{sec:nonthermal}

In Section~\ref{sec:Rovib_temps}, we have shown that the BN/KL outflow appears to be well-characterized by a thermal distribution. The high velocities ($\textgreater$40 km s$^{-1}$) of the \HH~fingers and the presence of \FeII~emission \citep{Bally2015,Youngblood2016a} indicate that many of these shocks are likely dissociative, have high post-shock temperatures, and therefore emit UV radiation \citep{Wolfire1991}. UV continuum or emission line flux can create non-thermal populations via fluorescence. In order to better characterize the excitation conditions of the BN/KL outflow, we consider in this section signatures of \HH~fluorescence created specifically by \HI~\Lya~pumping (see \citealt{Shull1978} for an overview).

\subsection{General \Lya~pumping and radiative cascade} \label{sec:general_lya_pumping}

First, we outline the general process of \Lya~pumping and the resulting radiative cascade, and consider three test scenarios to find where the best signatures of \Lya~pumping exist. Similar radiative transfer models of \Lya~pumping of \HH~have been performed for protoplanetary disk and T Tauri stars showing signatures of \HH~absorption within observed \Lya~wings extending several hundred \kms~\citep{Herczeg2004,Schindhelm2012,Hoadley2017}.

A $\leq$200 km s$^{-1}$ wide \Lya~profile coincides spectrally with two strong Lyman band \HH~transitions. 1-2 R(6) (1215.73 \AA) and 1-2 P(5) (1216.07 \AA) pump \HH~molecules from the ground electronic state $X$ $^1$$\Sigma^+_g$ ($v_l$, $J_l$) = (2,6), (2,5), respectively, into the first excited electronic state $B$ $^1$$\Sigma^+_u$ ($v_u$, $J_u$) = (1,7), (1,4), respectively. The lifetimes of the (1,7) and (1,4) states are short ($\sim$10$^{-8}$ s; \citealt{Abgrall1993}), and they decay back to an array of ($v_l$, $J_l$) levels in the ground electronic state following the dipole selection rules ($\Delta v$ = any, $\Delta J$ = $\pm$1). We refer to these as the (1,7) and (1,4) progressions, and the strongest of these transitions ($A_{ul}$~$\textgreater$ 5$\times$10$^7$ s$^{-1}$) are listed in Table~\ref{table:H2_electronic_lines}. Once back in the ground state, the (1,7) and (1,4) progressions cascade independently down to ($v_l$, $J_l$) = (0,0) and (0,1), respectively, following quadrupole selection rules ($\Delta v$ = any, $\Delta J$ = 0,$\pm$2, $J$ = 0 $\rightarrow$~0 forbidden). We treat the para and ortho states as independent species. 

\begin{figure}
   \begin{center}
   
     \subfigure{
          \includegraphics[width=\textwidth]{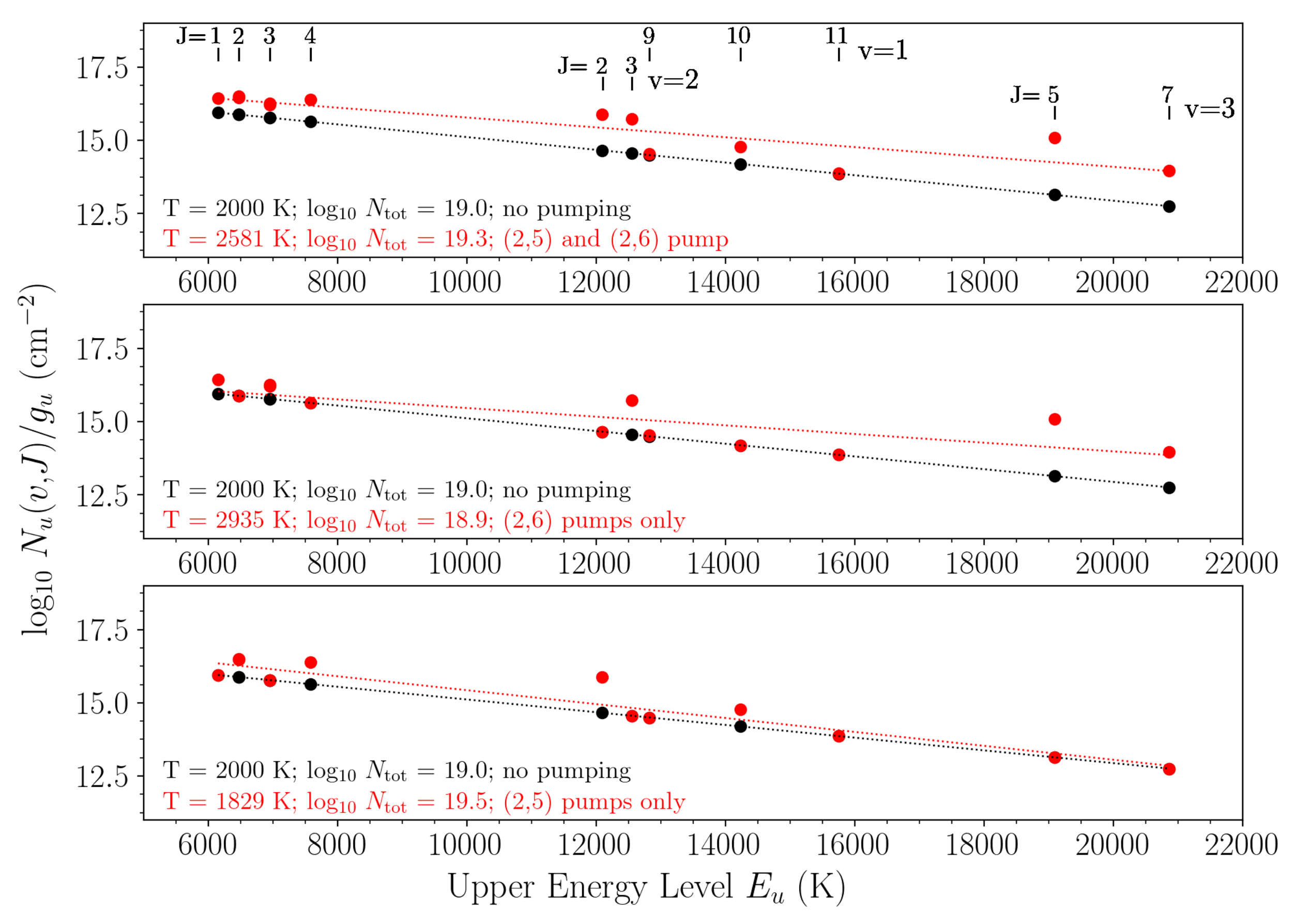}
          }

   \end{center}
    \caption{Example excitation diagrams using the 13 \HH~lines (Table~\ref{table:H2_lines}) showing purely thermal populations in black ($N$(\HH) = 10$^{18}$ cm$^{-2}$, $T$ = 2000 K, and ortho/para = 3), and excess population due to \Lya~pumping from ($v$,$J$) = (2,5) and (2,6) in red. The dotted black and red lines show the fits to the black and red data points, respectively, and the black and red text in the lower left corner of each subplot shows the fitted temperature and total \HH~column density corresponding to the color of the data points. \textit{Top:} (2,5) and (2,6) both absorbed 0.05 erg cm$^{-2}$ s$^{-1}$ sr$^{-1}$, selected as an illustrative amount. \textit{Middle:} (2,5) absorbed no flux, and (2,6) absorbed 0.05 erg cm$^{-2}$ s$^{-1}$ sr$^{-1}$. \textit{Bottom:} (2,5) absorbed 0.05 erg cm$^{-2}$ s$^{-1}$ sr$^{-1}$, and (2,6) absorbed no flux.
        }
    \label{fig:example_pumped_excitation_diagrams}

\end{figure}

We calculate the \Lya~flux pumped in the 1-2 R(6) and 1-2 P(5) lines using a procedure described in detail in \cite{McJunkin2016}. We calculate the optical depths ($\tau_{\lambda}$) of the Lyman band lines from absorption cross-sections ($\sigma_{lu}$) and column densities in the lower states $N$($v_l$,$J_l$). We then correct the optical depths as described in \cite{Liu1996} and \cite{Wolven1997}, because the wings of the two absorbing lines overlap. We attenuate the assumed intrinsic \Lya~profile using the corrected optical depths and integrate over the absorbed flux to calculate the pumping flux. The pumping fluxes are put directly into the (1,7) and (1,4) states of the first excited electronic state, where they are promptly redistributed via branching ratios ($r_{\rm branch}$ = $A_{ul}$/$\sum A_{ul}$; Table~\ref{table:H2_electronic_lines}) back into the ground electronic state. The probability of dissociation from the (1,7) and (1,4) upper levels is zero \citep{Abgrall2000,Herczeg2006}. 

We then compute the flux in each rovibrational ground state transition as the \HH~molecules cascade down, following the quadrupole selection rules of the ground electronic state ($\Delta v$ = any, $\Delta J$ = 0,$\pm$2, $J$ = 0 $\rightarrow$~0 forbidden). The cascade is computed using branching ratios calculated from the Einstein A coefficients from \cite{Wolniewicz1998}. We consider the cascade fluxes as an ``excess" compared to the fluxes determined from a thermally-populated ground state in equilibrium. Because the lifetimes of these ground state levels are long ($\sim$10$^7$ s; \citealt{Wolniewicz1998}), we assume that each cascading molecule that reaches ($v_l$,$J_l$) = (2,6) or (2,5) is re-pumped immediately by \Lya~into the (1,7) and (1,4) levels in the excited state. Effectively, the cascade ends when reaching the (2,6) or (2,5) levels, where it is re-pumped into (1,7) and (1,4). This re-pumped flux is 15.3\% of the initial pumping flux for the (1,7) progression and 2.5\% for the (1,4) progression. Allowing multiple pumpings creates a small effect on the resulting excitation diagrams as the fitted temperatures are changed by $\leq$100 K and the column densities are changed by $\leq$0.1 dex. We find that more than three multiple pumpings causes negligible additional modification of the \HH~populations, and we truncated the multiple pumping chain at three.  

Figure~\ref{fig:example_pumped_excitation_diagrams} shows an example of the excess flux in our 13 \HH~lines produced by the radiative cascade following \Lya~pumping. We consider three limiting cases where (a) the \Lya~flux pumped into (1,7) and (1,4) are equal, (b) flux was pumped into (1,7) but not (1,4), and (c) flux was pumped into (1,4) but not (1,7). 0.05 erg cm$^{-2}$ s$^{-1}$ sr$^{-1}$ was chosen as an example value that would have a large effect on the excitation diagram. Of our 13 near-IR \HH~lines (Table~\ref{table:H2_lines}), we find that all but 1-0 S(7), 1-0 S(8), and 1-0 S(9) should show excess flux due to \Lya~pumping if both the (2,6) and (2,5) states are pumped. We also consider other near-IR and mid-IR \HH~lines, many of which have been detected in Orion BN/KL by other instruments \citep{Rosenthal2000,Oh2016,Geballe2017} or are expected to be detected. The lines that show the largest increases are the 0-0 S-branch lines in the mid-IR from 8--28.2 $\mu$m. However, excess flux in low ($v$,$J$) emission lines will be greatly diminished by collisional de-excitation, which is discussed more in Section~\ref{sec:caveats}. In Table~\ref{table:otherlines}, we present a list of \HH~transitions through which $\geq$0.1\% of the pumping \Lya~flux passes during the radiative cascade. In general, lines originating from $J_u$~$\gtrsim$~10 receive negligible \Lya~flux.

\subsection{Intrinsic \Lya~from the post-shock region} \label{sec:intrinsic_lya}

We assume a \Lya~profile that originates from the \HH~fingers themselves. Fast, dissociative shocks (J-type shocks) emit UV radiation, including \Lya~that has been observed to pump rovibrationally excited \HH~populations (e.g., \citealt{Wolfire1991}). \cite{Walter2003} found evidence of this in the outflow from T Tauri; the \Lya~radiation is described by a profile that is narrow compared to the broad (several hundred km s$^{-1}$) \Lya~emission from the accreting star. \cite{Walter2003} do not place a direct constraint on the line shape, but find that only the 1-2 R(6) and 1-2 P(5) Lyman band \HH~transitions are responsible for the observed fluorescence in the outflow. The velocity separation of these two transitions, 1-2 R(6) at 1215.73 \AA~and 1-2 P(5) at 1216.07 \AA, indicates that the \Lya~line must have significant flux over 0.34 \AA~(84 km s$^{-1}$), as measured at the base of the line.

We assume a narrow (40 \kms) and a broad (100 \kms) profile. From the shock models of \cite{Hartigan1987}, we estimate the \Lya~integrated flux through the front of the shock to be approximately 10$^{-3}$ erg cm$^{-2}$ s$^{-1}$ sr$^{-1}$ for a variety of shock velocities. This model surface brightness is in agreement with a simple calculation that assumes every H atom passing through the shock is ionized, then recombines. 68\% of recombinations under Case B assumptions (optically thick in all Lyman lines) result in a \Lya~photon, so the \Lya~photon surface brightness through the shock is 1/(4$\pi$)$\cdot$0.68$\cdot$$n_{\rm pre-shock}$$\cdot$$V_{\rm shock}$, where $n_{\rm pre-shock}$ is the pre-shock hydrogen density, and $V_{\rm shock}$ is the shock velocity. Assuming $n_{\rm pre-shock}$ = 100 cm$^{-2}$ and $V_{\rm shock}$ = 100 km s$^{-1}$ (typical values for the region; \citealt{Bally2015}), we find a flux of 8.8$\times$10$^{-4}$ erg cm$^{-2}$ s$^{-1}$ sr$^{-1}$. As these two values are in good agreement, we assume 10$^{-3}$ erg cm$^{-2}$ s$^{-1}$ sr$^{-1}$ as the \Lya~surface brightness passing through the shock.

We assume no self-reversal and that the \Lya~emission and \HH~absorption have velocity centroids of 0 km s$^{-1}$. For the \HH~population, we assume $T$ = 2000 K, $N$(\HH) = 10$^{19}$ cm$^{-2}$ and a Doppler broadening value $b$ = 5 \kms~(includes a turbulent velocity of 2 \kms). For a narrow profile (40 \kms), 4$\times$10$^{-4}$ erg cm$^{-2}$ s$^{-1}$ sr$^{-1}$ is pumped into (1,7) and no flux is pumped into (1,4). For a broad profile (100 \kms), 1.7$\times$10$^{-4}$ erg cm$^{-2}$ s$^{-1}$ sr$^{-1}$ is pumped into (1,7), and 3.6$\times$10$^{-5}$ erg cm$^{-2}$ s$^{-1}$ sr$^{-1}$ is pumped into (1,4). In both cases, the \Lya~enhancements in the near-IR lines are negligible and would not be detected by TripleSpec. The derived temperatures from the resulting excitation diagram are within 0.5\% of the thermal population's temperature. Unless the \Lya~surface brightness estimate is severely underestimated, \Lya~pumping signatures from shocks should not be detectable in the near-IR. In the mid-IR, the 0-0 S(0), 0-0 S(1), and 0-0 S(2) show strong enhancements above their thermal flux values, but these lines should be subject to rapid re-thermalization due to collisional de-excitation with H atoms.

\subsection{Nebular \Lya} \label{sec:nebular_lya}

We consider the possibility that resonant \Lya~scattering from the bright \HII~region (M42) in the foreground could excite the \HH~fingers, akin to the planetary nebula scenario in \cite{Lupu2006}. The source of the ionizing photons that create the \HII~region is the Trapezium cluster, and the brightest member, $\theta^1$~Ori C, is the dominant driver of the \HII~region. The Trapezium cluster is approximately 0.25 pc in front of the ionization front separating the foreground \HII~region and the background OMC1 cloud core from which the BN/KL outflow emerges \citep{Wen1995,Odell2001}. Nebular \Lya, created from \HI~recombination, resonantly scatters around the nebula, impinging on the interface between the \HII~region and the molecular cloud. Any \Lya~that escapes will be quickly attenuated over the sharp increase in density across the interface region and will not penetrate to most of the BN/KL outflow. However, some of the \HH~fingers have low extinction values and visibility at optical wavelengths indicate that they are poking through the ionization front \citep{Graham2003}. 

We determine the average \Lya~surface brightness of the nebula from H$\alpha$~(6563 \AA). From the VLT/MUSE H$\alpha$ image of \cite{Weilbacher2015}, we measure the mean H$\alpha$~surface brightness across the nebula to be $F$(H$\alpha$) = 4.31$\times$10$^{-2}$ erg cm$^{-2}$ s$^{-1}$ sr$^{-1}$. Assuming Case A recombination as an upper limit, the flux ratio between \Lya~and H$\alpha$ $F$(\Lya)/$F$(H$\alpha$)~$\approx$~11 at $T$ = 10$^4$ K \citep{Hummer1987}. Thus, $F$(\Lya) = 0.47 erg cm$^{-2}$ s$^{-1}$ sr$^{-1}$.

To determine the amount of dust and gas in the interface region separating the nebular \Lya~photons and the \HH~fingers, we compare the Br$\gamma$/Pa$\beta$ extinction map that measures the dust foreground to the nebula (described in Section~\ref{sec:ObservationsReductions}) to the extinction map based on the flux ratio of \HH~1-0 Q(3) and 1-0 S(1) from \cite{Youngblood2016a}, which measures the dust foreground to the nebula and the dust between the \HH~fingers and the nebula. By subtracting the two extinction maps, we measure the dust extinction between the \Lya~photons and the BN/KL outflow, and find the lowest values to be $A_{\rm V}$~$\sim$~0.5--1 mag. Assuming $R_{\rm V}$ = 5.5, a typical value for star forming regions in the Milky Way with large dust grains, and extinction curves presented in \cite{Draine2011}, $A_{\rm V}$ = 1 corresponds to $A_{\rm Ly\alpha}$ = 1.4. Therefore, the nebular \Lya~is attenuated by at least 70\% due to dust alone ($A_{\lambda}$= 2.5~$\log_{10}$[$F_{\lambda}^{\rm emit}$/$F_{\lambda}^{\rm obs}$]). Assuming no \HH, the column of \HI~gas associated with $E$($B$--$V$) = $A_{\rm V}$/$R_{\rm V}$ = 0.182 is $N$(\HI) = 4.93$\times$10$^{21}$ $\times$~$E$($B$--$V$) cm$^{-2}$ mag$^{-1}$ = 9.0$\times$10$^{20}$ cm$^{-2}$ \citep{Diplas1994}. For this \HI~column density and Doppler broadening values $b$~$\leq$~100 \kms, the optical depth of the \Lya~line core is 10$^{6}$~$\leq$~$\tau_{\rm Ly\alpha}$~$\leq$~10$^8$ and the attenuation only becomes non-zero outside $\pm$500 \kms. There are other \HH~electronic transitions within $\pm$1000 \kms~of \Lya~that could be pumped (e.g., \citealt{Herczeg2006}), but nebular \Lya~emission would not be broad enough and/or Doppler-shifted enough to overlap with these transitions.

\subsection{Limitations of the \Lya~pumping model} \label{sec:caveats}

Our simple model neglects collisional de-excitation, or re-thermalization, of the level populations on timescales shorter than the radiative decay timescale. The critical density, where de-excitation due to collisions with H atoms is equal to the rate of radiative de-excitation, is $n$(\HI)~$\sim$~10$^6$ cm$^{-3}$ for lines like 1-0 S(1), 1-0 S(2), and 2-1 S(1) at 2000 K \citep{LeBourlot1999}. \cite{Bally2015} estimate that the densities of the \HH~bullets must be approximately 10$^4$ times greater than the medium through which they are traveling in order to have sustained their motion over large distances. The medium's density is 10$^2$--10$^3$ cm$^{-3}$, so the number density of the bullets is likely $n$(\HH)~$\sim$~10$^7$ cm$^{-3}$. Assuming $n$(\HI)/$n$(\HH)~$\sim$~10$^{-1}$, which is in between the typical values for shocks associated with low-mass star formation and PDRs \citep{LeBourlot1999}, $n$(\HI) is approximately the critical density. However, this estimate assumes that the entire bullet mass is shocked, which is likely not the case. We provide a second estimate of the shocked gas density by assuming that the \HH~emission spreads uniformly across the typical width of an emitting \HH~knot (4\arcsec~= 2.4$\times$10$^{16}$ cm) and the path length is equal to the width. Assuming a large total \HH~column density ($N$(\HH) = 10$^{20}$ cm$^{-2}$), we find that $n$(\HH)~$\sim$~4000 cm$^{-3}$ and therefore $n$(\HI)~$\sim$~400 cm$^{-3}$, indicating that the observed near-IR transitions would not be significantly affected by collisional de-excitation. Based on the second density estimate, we conclude that observed near-IR transitions are only affected to a small degree by collisional de-excitation. However, lower ($v$,$J$) transitions will be significantly affected, such as the 0-0 rotational lines which show large excess flux signatures in the radiative cascade calculation (Table~\ref{table:otherlines}).

The excess flux calculation does not account for the potentially lower-than-thermal fluxes in quadrupole lines originating from the (2,6) and (2,5) states. Given the long lifetimes of the (2,6) and (2,5) states and the strength of the two Lyman band transitions pumping out of this state, we do not allow radiation in these levels to continue to cascade down to the ground state. If it is physically accurate that all of the \HH~molecules from the (2,6) and (2,5) levels should be immediately pumped, then the observed column densities in these levels should be very small, and lower ($v$,$J$) levels should also be affected.

\begin{deluxetable}{clcc|clcc}
\tablecolumns{8}
\tablewidth{0pt}
\tablecaption{Electronic \HH~Transitions: the (1,4) and (1,7) progressions with $A_{ul}$ $\textgreater$~10$^7$ s$^{-1}$ \label{table:H2_electronic_lines}} 
\tablehead{\colhead{Transition} & 
                  \colhead{($v_u$,$J_u$)~$\rightarrow$} &
                  \colhead{$\lambda_0$ (\AA)} &
                  \colhead{$A_{ul}$} &
                  \colhead{Transition} & 
                  \colhead{($v_u$,$J_u$)~$\rightarrow$} &
                  \colhead{$\lambda_0$ (\AA)} &
                  \colhead{$A_{ul}$} \\
                  \colhead{} &
                  \colhead{($v_l$,$J_l$)} &
                  \colhead{} &
                  \colhead{(10$^{7}$ s$^{-1}$)} &
                  \colhead{} &
                  \colhead{($v_l$,$J_l$)} &
                  \colhead{} &
                  \colhead{(10$^{7}$ s$^{-1}$)}
                  }
\startdata
1-1 R(3) & (1,4)$\rightarrow$(1,3) & 1148.70 & 6.70 & 1-1 R(6) & (1,7)$\rightarrow$(1,6) & 1161.95 & 6.35\\
1-1 P(5) & (1,4)$\rightarrow$(1,5) & 1161.82 & 8.47 & 1-1 P(8) &  (1,7)$\rightarrow$(1,8) & 1183.31 & 7.53 \\
1-2 R(3) & (1,4)$\rightarrow$(2,3) & 1202.45 & 13.38 & 1-2 R(6)* & (1,7)$\rightarrow$(2,6) & 1215.73 & 13.62\\
1-2 P(5)* & (1,4)$\rightarrow$(2,5) & 1216.07 & 15.92 & 1-2 P(8) & (1,7)$\rightarrow$(2,8) & 1237.87 & 14.69 \\
1-3 R(3) & (1,4)$\rightarrow$(3,3) & 1257.83 & 11.92 & 1-3 R(6) & (1,7)$\rightarrow$(3,6) & 1271.02 & 13.42\\
1-3 P(5) & (1,4)$\rightarrow$(3,5) & 1271.93 & 12.76 & 1-3 P(8) & (1,7)$\rightarrow$(3,8) & 1293.87 & 12.27\\
1-6 R(3) & (1,4)$\rightarrow$(6,3) & 1431.01 & 9.98 & 1-6 R(6) & (1,7)$\rightarrow$(6,6) & 1442.87 & 9.32\\
1-6 P(5) & (1,4)$\rightarrow$(6,5) & 1446.12 & 14.19 & 1-6 P(8) & (1,7)$\rightarrow$(6,8) & 1467.08 & 13.46\\
1-7 R(3) & (1,4)$\rightarrow$(7,3) & 1489.57 & 16.24 & 1-7 R(6) & (1,7)$\rightarrow$(7,6) & 1500.45 & 16.97\\
1-7 P(5) & (1,4)$\rightarrow$(7,5) & 1504.76 & 19.76 & 1-7 P(8) & (1,7)$\rightarrow$(7,8) & 1524.65 & 18.70\\
1-8 R(3) & (1,4)$\rightarrow$(8,3) & 1547.34 & 11.48 & 1-8 R(6) & (1,7)$\rightarrow$(8,6) & 1556.87 & 12.50\\
1-8 P(5) & (1,4)$\rightarrow$(8,5) & 1562.39 & 12.27 & 1-8 P(8) & (1,7)$\rightarrow$(8,8) & 1580.67 & 10.99\\
\enddata
\tablecomments{All transitions are between $B$ $^1$$\Sigma^+_u$ and $X$ $^1$$\Sigma^+_g$. Wavelengths and Einstein A coefficients from \cite{Abgrall1993}.}
\tablenotetext{*}{1-2 P(5) and 1-2 R(6) are the transitions pumped by \Lya.}
\end{deluxetable}

\startlongtable
\begin{deluxetable}{ccccc|ccccc}
\tablecolumns{10}
\tablewidth{0pt}
\tabletypesize{\scriptsize}
\tablecaption{Rovibrational \HH~transitions receiving $\geq$0.1\% of \Lya~pumping flux \label{table:otherlines}} 
\tablehead{\colhead{Transition} & 
                  \colhead{$\lambda_0$ (\AA)} &
                  \colhead{$E_u$ (K)} &
                  \colhead{$A_{ul}$} &
                  \colhead{\Lya~cascade$^{\dagger}$} &
                  \colhead{Transition} & 
                  \colhead{$\lambda_0$ (\AA)} &
                  \colhead{$E_u$ (K)} &
                  \colhead{$A_{ul}$} &
                  \colhead{\Lya~cascade$^{\dagger}$} \\
                  \colhead{} &
                  \colhead{} &
                  \colhead{} &
                  \colhead{(10$^{-7}$ s$^{-1}$)} &
                  \colhead{(\%)} &
                  \colhead{} &
                  \colhead{} &
                  \colhead{} &
                  \colhead{(10$^{-7}$ s$^{-1}$)} &
                  \colhead{(\%)}
                  }
\startdata
0-0 S(8) & 50376 & 8687 & 3.23 & 1.2 & 3-2 Q(2) & 27186 & 17387 & 4.84 & 0.85 \\
0-0 S(7) & 55025 & 7201 & 2.0 & 0.20 & 3-2 Q(3) & 27312 & 17817 & 4.42 & 3.6 \\
0-0 S(6) & 61041 & 5831 & 1.14 & 10.6 & 3-2 Q(4) & 27481 & 18385 & 4.18 & 2.2 \\
0-0 S(5) & 69074 & 4586 & 0.59 & 3.3 & 3-2 Q(5) & 27693 & 19085 & 3.99 & 2.3 \\
0-0 S(4) & 80241 & 3474 & 0.26 & 35.7 & 3-2 Q(6) & 27948 & 19912 & 3.79 & 4.1 \\
0-0 S(3) & 96645 & 2503 & 0.1 & 18.4 & 3-2 Q(7) & 28249 & 20857 & 3.58 & 0.22 \\
0-0 S(2) & 122784 & 1681 & 0.03 & 63.9 & 3-2 O(2) & 29620 & 16952 & 14.1 & 0.11 \\
0-0 S(1) & 170346 & 1015 & 0.005 & 50.0 & 3-2 O(3) & 31638 & 17097 & 6.87 & 2.2 \\
0-0 S(0) & 282184 & 509 & 0.0003 & 84.8 & 3-2 O(4) & 33959 & 17387 & 4.87 & 0.86 \\
1-1 S(8) & 53100 & 14233 & 2.92 & 0.49 & 3-2 O(5) & 36635 & 17817 & 3.52 & 2.9 \\
1-1 S(6) & 64315 & 11523 & 1.05 & 2.1 & 3-2 O(6) & 39727 & 18385 & 2.53 & 1.3 \\
1-1 S(5) & 72764 & 10342 & 0.54 & 0.11 & 3-2 O(7) & 43317 & 19085 & 1.79 & 1.1 \\
1-1 S(4) & 84511 & 9286 & 0.25 & 0.85 & 3-1 S(0) & 12620 & 17387 & 3.19 & 0.56 \\
1-1 S(3) & 101769 & 8365 & 0.09 & 0.20 & 3-1 S(1) & 12330 & 17817 & 4.68 & 3.8 \\
1-0 S(8) & 17124 & 14233 & 2.34 & 0.40 & 3-1 S(2) & 12075 & 18385 & 5.76 & 3.0 \\
1-0 S(6) & 17874 & 11523 & 3.54 & 7.1 & 3-1 S(3) & 11856 & 19085 & 6.59 & 3.9 \\
1-0 S(5) & 18355 & 10342 & 3.95 & 0.77 & 3-1 O(3) & 14180 & 17097 & 4.99 & 1.6 \\
1-0 S(4) & 18918 & 9286 & 4.19 & 14.4 & 3-1 O(4) & 14677 & 17387 & 2.86 & 0.50 \\
1-0 S(3) & 19575 & 8365 & 4.21 & 9.3 & 3-1 O(5) & 15220 & 17817 & 1.98 & 1.6 \\
1-0 S(2) & 20337 & 7584 & 3.98 & 9.4 & 3-1 O(6) & 15812 & 18385 & 1.38 & 0.72 \\
1-0 S(1) & 21218 & 6951 & 3.47 & 11.6 & 3-1 O(7) & 16455 & 19085 & 0.96 & 0.56 \\
1-0 S(0) & 22232 & 6471 & 2.53 & 4.0 & 4-3 S(6) & 21441 & 26622 & 2.3 & 0.48 \\
1-0 Q(1) & 24065 & 6148 & 4.29 & 9.1 & 4-3 S(5) & 21997 & 25626 & 3.24 & 0.17 \\
1-0 Q(2) & 24134 & 6471 & 3.03 & 4.7 & 4-3 S(4) & 22658 & 24735 & 4.04 & 2.0 \\
1-0 Q(3) & 24237 & 6951 & 2.78 & 9.3 & 4-3 S(3) & 23435 & 23956 & 4.6 & 0.97 \\
1-0 Q(4) & 24374 & 7584 & 2.65 & 6.3 & 4-3 S(1) & 25404 & 22761 & 4.51 & 1.6 \\
1-0 Q(5) & 24547 & 8365 & 2.55 & 5.6 & 4-3 S(0) & 26631 & 22354 & 3.49 & 0.16 \\
1-0 Q(6) & 24754 & 9286 & 2.44 & 8.4 & 4-2 S(0) & 13422 & 22354 & 5.25 & 0.23 \\
1-0 Q(7) & 24997 & 10342 & 2.34 & 0.46 & 4-2 S(1) & 13113 & 22761 & 7.56 & 2.7 \\
1-0 Q(8) & 25276 & 11523 & 2.23 & 4.4 & 4-2 S(2) & 12843 & 23296 & 9.15 & 2.8 \\
1-0 Q(10) & 25944 & 14233 & 1.99 & 0.34 & 4-2 S(3) & 12613 & 23956 & 10.3 & 2.2 \\
1-0 O(8) & 41642 & 9286 & 0.74 & 2.5 & 4-2 O(3) & 15094 & 22081 & 7.69 & 1.5 \\
1-0 O(7) & 38080 & 8365 & 1.06 & 2.3 & 4-2 O(4) & 15631 & 22354 & 5.17 & 0.23 \\
1-0 O(6) & 35009 & 7584 & 1.5 & 3.5 & 4-2 O(5) & 16219 & 22761 & 3.67 & 1.3 \\
1-0 O(5) & 32350 & 6951 & 2.08 & 6.9 & 4-2 O(6) & 16862 & 23296 & 2.63 & 0.82 \\
1-0 O(4) & 30038 & 6471 & 2.9 & 4.5 & 4-2 O(7) & 17562 & 23956 & 1.88 & 0.40 \\
1-0 O(3) & 28025 & 6148 & 4.22 & 9.0 & 4-2 O(8) & 18323 & 24735 & 1.33 & 0.65 \\
1-0 O(2) & 26268 & 5986 & 8.54 & 2.3 & 4-2 O(10) & 20058 & 26622 & 0.64 & 0.13 \\
2-1 S(8) & 18153 & 19449 & 2.43 & 0.18 & 5-3 Q(1) & 14909 & 26747 & 11.6 & 1.2 \\
2-1 S(6) & 18938 & 16883 & 4.31 & 5.7 & 5-3 Q(3) & 15036 & 27386 & 7.63 & 1.6 \\
2-1 S(5) & 19444 & 15764 & 5.06 & 0.74 & 5-3 Q(4) & 15138 & 27889 & 7.36 & 1.3 \\
2-1 S(2) & 21541 & 13150 & 5.6 & 6.5 & 5-3 Q(5) & 15267 & 28509 & 7.17 & 0.98 \\
2-1 S(1) & 22477 & 12549 & 4.98 & 7.4 & 5-3 Q(6) & 15424 & 29240 & 7.0 & 2.0 \\
2-1 S(0) & 23556 & 12094 & 3.68 & 1.7 & 5-3 Q(7) & 15608 & 30075 & 6.82 & 0.24 \\
2-1 Q(1) & 25509 & 11788 & 6.37 & 4.6 & 5-3 Q(8) & 15821 & 31008 & 6.64 & 1.1 \\
2-1 Q(2) & 25585 & 12094 & 4.49 & 2.1 & 5-3 O(3) & 16111 & 26747 & 11.1 & 1.2 \\
2-1 Q(3) & 25698 & 12549 & 4.12 & 6.2 & 5-3 O(5) & 17333 & 27386 & 5.53 & 1.1 \\
2-1 Q(4) & 25849 & 13150 & 3.91 & 4.5 & 5-3 O(6) & 18035 & 27889 & 4.05 & 0.74 \\
2-1 Q(7) & 26535 & 15764 & 3.41 & 0.50 & 5-3 O(7) & 18803 & 28509 & 2.97 & 0.41 \\
2-1 Q(8) & 26843 & 16883 & 3.23 & 4.3 & 5-3 O(8) & 19642 & 29240 & 2.16 & 0.61 \\
2-1 Q(10) & 27583 & 19449 & 2.84 & 0.20 & 6-4 S(1) & 14958 & 31700 & 11.4 & 2.9 \\
2-1 O(2) & 27861 & 11635 & 3.47 & 0.20 & 6-4 S(2) & 14658 & 32170 & 13.3 & 0.70 \\
2-1 O(3) & 29740 & 11788 & 6.39 & 4.7 & 6-4 S(3) & 14405 & 32748 & 14.3 & 3.5 \\
2-1 O(4) & 31898 & 12094 & 4.4 & 2.1 & 6-4 Q(1) & 15947 & 31103 & 14.0 & 0.59 \\
2-1 O(5) & 34379 & 12549 & 3.17 & 4.7 & 6-4 Q(3) & 16095 & 31700 & 9.15 & 2.4 \\
2-1 O(6) & 37239 & 13150 & 2.28 & 2.6 & 6-4 Q(4) & 16214 & 32170 & 8.79 & 0.47 \\
2-0 S(5) & 10849 & 15764 & 3.28 & 0.48 & 6-4 Q(5) & 16364 & 32748 & 8.52 & 2.1 \\
2-0 S(2) & 11382 & 13150 & 2.38 & 2.8 & 6-4 Q(6) & 16547 & 33430 & 8.27 & 2.1 \\
2-0 S(1) & 11622 & 12549 & 1.9 & 2.8 & 6-4 Q(7) & 16762 & 34209 & 8.01 & 0.18 \\
2-0 S(0) & 11895 & 12094 & 1.27 & 0.60 & 6-4 Q(8) & 17013 & 35078 & 7.73 & 2.5 \\
2-0 Q(1) & 12383 & 11788 & 1.61 & 1.2 & 6-4 O(3) & 17246 & 31103 & 14.0 & 0.59 \\
2-0 Q(2) & 12419 & 12094 & 1.38 & 0.66 & 6-4 O(5) & 18580 & 31700 & 7.23 & 1.9 \\
2-0 Q(3) & 12473 & 12549 & 1.29 & 1.9 & 6-4 O(6) & 19352 & 32170 & 5.41 & 0.29 \\
2-0 Q(4) & 12545 & 13150 & 1.25 & 1.4 & 6-4 O(7) & 20200 & 32748 & 4.05 & 1.0 \\
2-0 O(2) & 12932 & 11635 & 12.8 & 0.73 & 7-5 O(5) & 19983 & 35709 & 8.45 & 1.4 \\
2-0 O(3) & 13354 & 11788 & 1.94 & 1.4 & 7-5 O(7) & 21780 & 36681 & 4.88 & 0.80 \\
2-0 O(4) & 13816 & 12094 & 1.03 & 0.49 & 7-5 Q(3) & 17287 & 35709 & 9.8 & 1.6 \\
2-0 O(5) & 14321 & 12549 & 0.7 & 1.0 & 7-5 Q(5) & 17606 & 36681 & 9.01 & 1.5 \\
2-0 O(6) & 14870 & 13150 & 0.47 & 0.55 & 7-5 Q(6) & 17823 & 37312 & 8.68 & 2.5 \\
3-2 S(6) & 20119 & 21915 & 3.59 & 2.5 & 7-5 Q(8) & 18381 & 38836 & 7.93 & 2.7 \\
3-2 S(5) & 20650 & 20857 & 4.53 & 0.27 & 7-5 S(1) & 16055 & 35709 & 11.7 & 1.9 \\
3-2 S(4) & 21278 & 19912 & 5.25 & 5.7 & 7-5 S(3) & 15479 & 36681 & 13.8 & 2.3 \\
3-2 S(3) & 22013 & 19085 & 5.65 & 3.3 & 7-5 S(4) & 15268 & 37312 & 13.6 & 3.9 \\
3-2 S(2) & 22870 & 18385 & 5.65 & 2.9 & 7-5 S(6) & 14993 & 38836 & 11.3 & 3.8 \\
3-2 S(1) & 23865 & 17817 & 5.15 & 4.2 & 8-6 O(5) & 21568 & 39419 & 8.81 & 1.3 \\
3-2 S(0) & 25014 & 17387 & 3.89 & 0.69 & 8-6 S(1) & 17301 & 39419 & 10.6 & 1.6 \\
3-2 Q(1) & 27102 & 17097 & 7.05 & 2.3 & 8-6 S(3) & 16707 & 40311 & 11.7 & 1.8 \\
\enddata
\tablecomments{Upper level energies ($E_{u}$) and wavelengths ($\lambda_0$) were calculated through \citealt{Herzberg1950}. Einstein A coefficients ($A_{ul}$) are from \citealt{Wolniewicz1998}.}
\tablenotetext{\dagger}{As in Table~\ref{table:H2_lines}, the \Lya~cascade column denotes what percentage of the flux pumped from the (2,5) and (2,6) levels cascades through these lines. The ortho transitions' percentages are calculated by normalizing to the flux pumped out of (2,5) and the para transitions' are calculated by normalizing to the flux pumped out of the (2,6) level of the ground state.}
\end{deluxetable}

\section{Summary} \label{sec:Summary}

Using the \HH~de-reddened integrated intensity maps from \cite{Youngblood2016a}, we have measured the \HH~temperature and column density on a spaxel-by-spaxel basis across the Orion BN/KL outflow. Most of the region is well-characterized by single-temperature fits with $T$(\HH)~$\sim$~2000--2500 K, with the \HH~fingers farthest from the outflow origin showing higher temperatures and lower column densities. This gas comprises 10$^{-5}$--10$^{-3}$ of the total \HH~column density. We also show that our set of 13 \HH~lines from the near-IR H and K bands is not conducive to detecting multiple temperature components, and our measured temperatures are subject to biases based on the range of the upper level energies probed by the lines. 

Comparing our column density results to the CO column density of $T$=2500 K gas measured by \textit{Herschel} \citep{Goicoechea2015}, we find CO/\HH~=~2$\times$10$^{-3}$, which is in significant excess of the canonical 2.7$\times$10$^{-4}$ value \citep{Lacy1994}. The CO column density was measured from lines near 100 \mum, and the \HH~column density was measured near 2 \mum, so dust extinction could  explain this difference. Other possible effects on the observed CO/\HH~fractional abundance could be incorrect assumptions on co-spatiality of the CO and \HH~and an overestimation of the CO column density. If the true CO/\HH~abundance is in agreement with the canonical value and only dust extinction is affecting our measurement of CO/\HH, then an extincting layer of $A_{\rm V}$~= 17 mag must lie behind the observed \HH, in addition to the foreground extinction of $A_{\rm V}$~$\sim$~5 mag. This would mean that we are only observing $\sim$10\% of the $T$=2500 K \HH~gas; the rest is completely obscured by dust.

We have also created a simple \Lya~pumping model for warm \HH~populations, and we tabulate which near-IR and mid-IR lines should show evidence of excess flux due to fluorescence from Lyman band transitions. We find that in the presence of strong \Lya~emission, fluorescence signatures are present in bright near-IR lines in the H and K bands, but we calculate that incident \Lya~radiation is likely very small and would result in extremely weak fluorescence signatures. We do not detect \Lya~fluorescence signatures in our excitation diagrams, and the 2000--2500 K temperatures of the region are consistent with shock heating.

\acknowledgments
We thank the referee for suggestions that improved the manuscript, and Jeremy Darling, Jason Glenn, Serena Criscuoli, and Mih\'aly Hor\'anyi for thoughtful guidance on the methodology and analysis.

\facilities{Apache Point Observatory (TripleSpec)}
\software{Astropy \citep{Robitaille2013}, IPython \citep{Perez2007}, Matplotlib \citep{Hunter2007}, molecular-hydrogen (https://github.com/keflavich/molecular\_hydrogen), NumPy and SciPy \citep{VanderWalt2011}.}

\end{document}